\documentclass{jfm}

\usepackage{caption}
\usepackage{graphicx}
\usepackage{natbib}
\usepackage{color}
\usepackage{amsmath}
\usepackage{epsfig}
\usepackage{lineno}
\usepackage{subfigure}
\usepackage{cases}
\usepackage[normalem]{ulem}
\usepackage{gensymb}

\DeclareFontFamily{OT1}{pzc}{}
\DeclareFontShape{OT1}{pzc}{m}{it}{<-> s * [1.10] pzcmi7t}{}
\DeclareMathAlphabet{\mathpzc}{OT1}{pzc}{m}{it}

\newcommand{\ve}{\varepsilon}
\newcommand{\eps}{\delta}

\newcommand{\bur}{\bar{u}_{r}}

\newcommand{\mcalM}{\mathcal{G}}

\ifCUPmtlplainloaded \else
  \checkfont{eurm10}
  \iffontfound
    \IfFileExists{upmath.sty}
      {\typeout{^^JFound AMS Euler Roman fonts on the system,
                   using the 'upmath' package.^^J}%
       \usepackage{upmath}}
      {\typeout{^^JFound AMS Euler Roman fonts on the system, but you
                   dont seem to have the}%
       \typeout{'upmath' package installed. JFM.cls can take advantage
                 of these fonts,^^Jif you use 'upmath' package.^^J}%
      }
  \else
  \fi
\fi

\ifCUPmtlplainloaded \else
  \checkfont{msam10}
  \iffontfound
    \IfFileExists{amssymb.sty}
      {\typeout{^^JFound AMS Symbol fonts on the system, using the
                'amssymb' package.^^J}%
       \usepackage{amssymb}%

      }{}
  \fi
\fi

\ifCUPmtlplainloaded \else
  \IfFileExists{amsbsy.sty}
    {\typeout{^^JFound the 'amsbsy' package on the system, using it.^^J}%
     \usepackage{amsbsy}}
    {\providecommand\boldsymbol[1]{\mbox{\boldmath $##1$}}}
\fi

\newcommand\Pe{\mbox{\textit{Pe}}}  % Peclet number
\newsavebox{\astrutbox}
\sbox{\astrutbox}{\rule[-5pt]{0pt}{20pt}}

\def\XXint#1#2#3{{\setbox0=\hbox{$#1{#2#3}{\int}$}
     \vcenter{\hbox{$#2#3$}}\kern-.5\wd0}}

\newcommand\Ca{\mbox{\textit{Ca}}}

\title[The nascent coffee ring: how solute diffusion counters advection]{The nascent coffee ring: how solute diffusion counters advection}

\author[M. R. Moore, D. Vella \& J. M. Oliver]%
{M.\ns R.\ns M\ls O\ls O\ls R\ls E$^1$, \ns 
D.\ns V\ls E\ls L\ls L\ls A$^1$ \ns
\and
J.\ns M.\ns O\ls L\ls I\ls V\ls E\ls R$^1$}

\affiliation{$^1$Mathematical Institute, University of Oxford, Andrew Wiles Building, Radcliffe Observatory Quarter, Woodstock Road, Oxford, OX2 6GG}

\pubyear{}
\volume{}
\pagerange{}
\date{?; revised ?; accepted ?. - To be entered by editorial office}
\begin{document}

\maketitle

%%%%%%%%%%%%%%%%%%%%%%%%%%%%%%%%%%%%%%%%%%%%%%%%%%%%%%%%%%%%%%%
%%%%%%%%%%%%%%%%%%%%%%%%%%% ABSTRACT %%%%%%%%%%%%%%%%%%%%%%%%%%
%%%%%%%%%%%%%%%%%%%%%%%%%%%%%%%%%%%%%%%%%%%%%%%%%%%%%%%%%%%%%%%

% \linenumbers

\begin{abstract}
We study the initial evolution of the coffee ring that is formed by the evaporation of a thin, axisymmetric, surface tension-dominated droplet containing a dilute solute. When the solutal P\'{e}clet number is large, we show that diffusion close to the droplet contact line controls the coffee-ring structure in the initial stages of evaporation. We perform a systematic matched asymptotic analysis for two evaporation models --- a simple, non-equilibrium, one-sided model (in which the evaporative flux is taken to be constant across the droplet surface) and a vapour-diffusion limited model (in which the evaporative flux is singular at the contact line) --- valid during the early stages in which the solute remains dilute. We call this the `nascent coffee ring' and describe the evolution of its features, including the size and location of the peak concentration and a measure of the width of the ring. Moreover, we use the asymptotic results to investigate when the assumption of a dilute solute breaks down and the effects of finite particle size and jamming are expected to become important. In particular, we illustrate the limited validity of this model in the diffusive evaporative flux regime.
\end{abstract}

\section{Introduction}

A droplet of coffee left to evaporate into the surrounding air leaves behind a 
stain that is darkest towards its edge, a phenomenon known as the `coffee ring effect'. While there are variations depending on the particular properties of the liquid and solute under consideration, as well as the dominant mode of evaporation, the fundamental mechanism for the coffee ring is as follows. For many substrates, the droplet contact line becomes pinned by surface roughness or inhomogeneities. As the volatile liquid evaporates, an outward radial flow develops to replace the fluid evaporating from the pinned contact line \cite[][]{Deegan1997, Deegan2000}. This outward radial flow carries solute along with it. As further fluid is lost, this solute build up at the contact line eventually reaches its packing density, forming the coffee ring \cite[][]{Popov2005}. This phenomenon is not just restricted to coffee and is ubiquitous in situations involving liquids carrying a solute. It has even been shown to be possible in initially-pure liquid droplets that evaporate on a substrate that dissolves on a faster timescale than the evaporative process \cite[][]{Mailleur2018}. Depending on the physical situation, the coffee ring effect may be advantageous. For example, the outward flow that drives the effect can be used to align DNA to aid mapping \cite[][]{Jing1998, Smalyukh2006}, to order arrays of nanoscopic structures \cite[][]{Kimura2003} or colloids \cite[][]{Koh2006}, or to aid the patterning of colloidal films \cite[][]{Harris2007}. However, in other situations, the tendency of this flow to produce an inhomogenous deposit may be undesirable. Examples include when one requires a uniform deposit in dip-coating \cite[][]{Berteloot2008} or in the formation of cDNA microarrays \cite[][]{Blossey2002,Blossey2003}.

Given the ubiquity of volatile liquid droplets containing a solute, an understanding of the physical mechanisms behind the coffee ring effect has been of great interest to researchers over the recent decades. The seminal work of \citet{Deegan1997} and \citet{Deegan2000} first linked the appearance of the coffee ring to the flow induced by evaporation. Since Deegan and coworkers assumed that the evaporation of the droplet is dominated by diffusive processes in the vapour, the evaporative flux is singular at the pinned edge of the droplet and so, moving to preserve conservation of mass, an outward flow develops in the droplet, taking fluid and solute to the contact line. \citet{Deegan1997} and \citet{Deegan2000} developed an analytical model for an axisymmetric droplet, deriving an expression for the amount of solute mass transported to the contact line as a function of the drying time. They show that the stagnation point flow of the droplet drives all the solute mass to the contact line by the time the droplet has completely evaporated. This mass is concentrated into a ring of infinitesimal width at the contact line.

For droplets with larger initial contact angles, \citet{Kang2016} describe an alternative mechanism for solute transport, in which the solute particles are captured by the rapidly diminishing droplet free surface and subsequently transported along the free surface to the pinned contact line. However this process is dominated by the radial capillary flow in the thin-drop (vanishing contact angle) limit.

\citet{Hu2002} extended the analysis of \citet{Deegan2000} to consider the role of the droplet contact angle in determining the diffusive evaporative flux from the surface of the droplet, concluding that it plays an important role only for droplets whose (macroscopic) contact angle is larger than 40$\degree$. In the limit of thin droplets, the diffusive evaporative flux is well approximated by that for a flat disk of liquid. 

Of course, no coffee ring is really completely located at the contact line --- the coffee ring must in fact have a concentration profile. \citet{Kajiya2008} performed a number of fluorescent microscopy experiments that show how the coffee ring varies for two different evaporative flux laws. They show that for droplets that are confined within a box --- which restricts how the vapour concentration can move away from the droplet --- the evaporation rate is essentially uniform across the droplet. Moreover, although the coffee ring effect is still observed, the thickness of the ring is much larger in the constant-evaporation case compared to droplets that are allowed to evaporate naturally into the surrounding gas, i.e. diffusively. It is this idea of using geometry to alter the evaporative flux of the droplet that is used in, for example, the patterning techniques of \citet{Harris2007}.

The evaporative flux can also be manipulated by changing the surrounding environment. \citet{Boulogne2016} compare the evaporative fluxes for an evaporating water droplet sitting on a dry substrate to a droplet sitting within a large hydrogel bath. In the latter case, the singular diffusive evaporative flux at the pinned contact line is greatly diminished by the hydrogel, which alters the vapour concentration in the surrounding gas. Even though there are weak convective effects, the evaporative flux can be well-approximated as a constant, and the authors demonstrate that, despite the change in flux, a coffee-ring still forms.

The pinned contact line plays a crucial role in the formation of a coffee ring. Indeed, if pinning can be inhibited by removing surface roughnesses \cite[][]{Marin2012} or coating the substrate in a hydrophilic oil \cite[][]{Li2020}, the coffee ring effect can be suppressed. Moreover, in cases where the contact line undergoes a stick-slip motion, multiple rings can form, see for example \citet{Adachi1995} and \citet{Shmuylovich2002}.

While a large amount of the literature has concentrated on the problem of an 
axisymmetric droplet due to its physical significance, there have been several recent studies analysing the effect of more general droplet profiles, including \citet{Witten2009} and \citet{Zheng2009}. In particular, \citet{Witten2009} derives a power law profile for the late-time deposit density, showing that it depends on the ratio of the evaporative flux above the stagnation point within the drop to the average evaporative flux over the whole droplet.
\citet{FreedBrown2015} and \citet{Saenz2017} have investigated the role of 
variable contact line curvature on the coffee ring effect, revealing that it is enhanced towards highly-curved parts of the contact line for a variety of different evaporative fluxes. 

An outstanding issue in the current analyses of the transfer of solute is 
that, while solute is transported advectively to the contact line, the mass at the contact line itself must vanish as the thickness of the droplet 
vanishes there. One way to address this deficiency is to consider the effect of the 
particle size in the model. In the model of \citet{Deegan1997}, the solute particle size is assumed to have no effect on the flow dynamics in the bulk --- an assumption that breaks down close to the contact line, where the concentration increases as the droplet evaporates. When the solute concentration reaches a sufficient level, the size of the particles has a leading-order effect on the local flow, so that two-phase suspension models are appropriate to describe the dynamics (see \citet{Guazzelli2018} and references therein). Furthermore, eventually the particle packing density is reached, leading to the possibility of the solute jamming close to the contact line. Such a model is considered by \citet{Popov2005}, who allows the solute to effectively jam within the fluid when the local solute concentration reaches a threshold 
value. When this occurs, no further solute can be transported into this region 
and is instead deposited sooner, leading to a thickening of the ring back towards the centre of the drop. \citet{Kaplan2015} extend this idea by considering the growing jammed region near the contact line as a porous medium, which in turn changes the local evaporative model. \citet{Kaplan2015} were able to show a transition from ring deposits to uniform deposits depending on the size of the capillary number and the initial solute concentration.

Jamming effects only come into play when the local solute concentration approaches the particle packing fraction. However, even before the packing fraction is reached, the evaporation-induced flow causes spatial gradients in the concentration, which must be resisted by diffusion (see figure \ref{fig:DropletConfigAxi}). The effect of diffusion in countering advection of solute is often neglected for one of two reasons. Firstly, the pertinent physical effect of interest is usually how much mass is transferred to which part of the boundary of the droplet. Secondly, the matched asymptotic analysis required to resolve the boundary layer in which diffusion matters is not straightforward, even in the limit in which the droplet is very thin. Our aim in this paper is to address this deficiency in the literature by considering a detailed matched asymptotic analysis for the solute transfer in an evaporating, pinned, axisymmetric droplet. In particular, we will describe the asymptotic structure and the resulting properties of the nascent coffee ring for two evaporation models: a kinetic evaporation model, in which the flux is taken to be constant, and a diffusive model, in which the flux is singular. We discuss the predictions of this model for the appearance of the characteristic `coffee ring' shape in the concentration profile well before jamming occurs, giving insight that may approximate the coffee ring height and thickness while the solute is still dilute. Crucially, our asymptotic predictions can also be used to assess the applicability of the dilute model and we demonstrate the particular importance of considering finite-particle-size effects in the diffusive evaporative flux regime.

%%%%%%%%%%%%%%%%%%%%%%%%%%%%%%%%%%%%%%%%%%%%%%%%%%%%%%%%%%%%%%%
%%%%%%%%%%%%%%%%%%%%% PROBLEM CONFIGURATION %%%%%%%%%%%%%%%%%%%
%%%%%%%%%%%%%%%%%%%%%%%%%%%%%%%%%%%%%%%%%%%%%%%%%%%%%%%%%%%%%%%

\section{Formulation of the mathematical model} \label{sec:ProblemConfig}

\subsection{The dimensional problem}

\begin{figure}
\centering \scalebox{1}{\epsfig{file=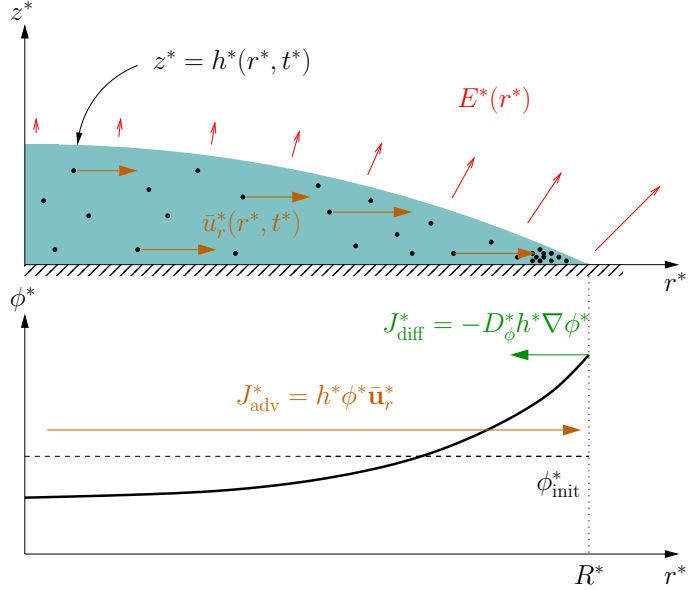}}
\caption{Schematic view of an axisymmetric liquid droplet evaporating on a substrate. The droplet footprint has radius $R^{*}$ and the liquid--air interface is denoted by $z^{*} = h^{*}(r^{*}, t^{*})$. As the droplet evaporates, the contact line remains pinned. Thus, to replace the mass lost to evaporation, an internal flow towards the contact line develops, which advects solute to the contact line with flux $J^{*}_{\mathrm{adv}}$. As the solute concentration $\phi^{*}$ increases close to the contact line, this induces a competing diffusive flux $J^{*}_{\mathrm{diff}}$ that opposes solute advection. This interplay leads to the formation of the nascent coffee ring.}
\label{fig:DropletConfigAxi} 
\end{figure}

We consider the configuration in figure \ref{fig:DropletConfigAxi} in which a droplet of liquid of volume $V^{*}$ lies on a rigid, planar substrate. The substrate lies along $z^{*} = 0$, where $(r^{*},\theta,z^{*})$ are cylindrical polar coordinates defined with respect to the centre of the droplet footprint, which is taken to be a circle of radius $R^{*}$. Here and hereafter, an asterisk denotes a dimensional variable. The droplet contact line is assumed to remain \emph{pinned} throughout the motion, which is a reasonable assumption for the majority of the drying time \cite[see][]{Hu2002} and certainly while the solute remains dilute. 

Following the symmetry of the problem, we make the assumption that the dynamics are independent of the polar angle $\theta$. The free surface delineating the droplet from the surrounding air is thus denoted by $z^{*} = h^{*}(r^{*},t^{*})$. We shall make the assumption that the droplet is \textit{thin} so that $H^{*} = h^{*}(0,0) \ll R^{*}$.

The liquid contains a non-volatile solute of initial concentration $\phi_{\mathrm{init}}^{*}$, which we shall assume to be evenly distributed throughout the droplet. We also assume throughout that the solute is sufficiently dilute that the flow within the drop is unaffected by its presence. This means that, crucially, we can decouple the flow in the liquid drop from solute transport.

\subsubsection{Flow problem}

The liquid has density $\rho^{*}$ and viscosity $\mu^{*}$, while the surface 
tension of the air-liquid interface is denoted by $\sigma^{*}$; all of these material parameters are taken to be constant. For the purposes of our analysis, we shall neglect the effect of gravity; that is, we assume the Bond number $\mbox{Bo} = \rho^{*} g^{*} R^{*2}/\sigma^{*}$ is small, where $g^{*}$ is the gravitational acceleration. The liquid velocity and pressure are denoted by $\textbf{u}^{*} = u_{r}^{*}\textbf{e}_{r} + u_{z}^{*}\textbf{e}_{z}$ and $p^{*}$ respectively, where $\textbf{e}_{r}, \textbf{e}_{z}$ are the unit vectors in the $r^{*}$- and $z^{*}$-directions.

Evaporation of the liquid into the surrounding air induces a flux of vapour $E^{*}$ at the droplet surface. We shall assume that the evaporation is a quasi-steady process, which is reasonable for a wide range of applications, including the evaporation of water on glass \cite[][]{Hu2002}. The evaporative flux in turn drives a flow within the droplet whose size $U^{*}$ depends upon the dominant evaporative process, as discussed in more detail in \textsection \ref{sec:NonDim}. 

In each of our evaporation models, the induced velocity will be assumed to be sufficiently small and the droplet sufficiently thin that, to leading-order in $\delta = H^{*}/R^{*}$, the equations of fluid motion within the drop are well-approximated by the lubrication equations \cite[][]{Deegan2000,FreedBrown2015}, i.e.
\refstepcounter{equation} 
\begin{linenomath}
$$
 \frac{\partial h^{*}}{\partial t^{*}} + \frac{1}{r^{*}}\frac{\partial}{\partial r^{*}}\left(r^{*}h^{*}\bar{u}_{r}^{*}\right) = -\frac{E^{*}}{\rho^{*}}, \quad  \bar{u}_{r}^{*} = -\frac{h^{*2}}{3\mu^{*}}\frac{\partial p^{*}}{\partial r^{*}}, \quad p^{*} = p_{\mathrm{atm}}^{*}-\frac{\sigma^{*}}{r^{*}}\frac{\partial}{\partial r^{*}}\left(r^{*}\frac{\partial h^{*}}{\partial r^{*}}\right)
 \eqno{(\theequation{\mathit{a},\mathit{b},\mathit{c}})}
 \label{eqn:FlowDim1}
$$ 
\end{linenomath}
for $0<r^{*}<R^{*}$, $t^{*}>0$, where $p_{\mathrm{atm}}^{*}$ denotes atmospheric pressure and $\bar{u}_{r}^{*}$ is the depth-averaged radial velocity. (Note that, in a slight abuse of language, we shall often refer to $\bur^{*}$ as simply the radial velocity for brevity.) Equations (\ref{eqn:FlowDim1}) must be solved subject to the symmetry conditions
\refstepcounter{equation} 
\begin{linenomath}
 $$
 \frac{\partial h^{*}}{\partial r^{*}} = r^{*}h^{*}\bar{u}_{r}^{*} = 0 \quad \mbox{at} \quad r^{*} = 0, \; t^{*}>0,
 \eqno{(\theequation{\mathit{a},\mathit{b}})}
 \label{eqn:FlowDim2}
 $$
\end{linenomath}
and the zero-thickness and no-flux conditions at the contact line
\refstepcounter{equation} 
\begin{linenomath}
 $$
 h^{*} = r^{*}h^{*}\bar{u}_{r}^{*} = 0 \quad \mbox{at} \quad r^{*} = R^{*} \; t^{*}>0.
 \eqno{(\theequation{\mathit{a},\mathit{b}})}
 \label{eqn:FlowDim3}
 $$
\end{linenomath}
While the initial droplet profile should capture its shape shortly after being deposited on the substrate and must be prescribed to close the problem (\ref{eqn:FlowDim1})--(\ref{eqn:FlowDim3}), after an initial transient on the timescale of capillary action (in which $t^{*} = \mu^{*}R^{*}/\delta^{3}\sigma^{*}$ by balancing the first two terms on the left-hand side of (\ref{eqn:FlowDim1}a)), the free surface rapidly approaches a spherical cap, see \citet{Lacey1982}. Hence, for simplicity, we shall impose the initial profile
\begin{linenomath}
 \begin{equation}
    h^{*}(r^{*},0) = 2V^{*}(R^{*2}-r^{*2})/\pi R^{*4} \quad \mbox{for} \quad 0<r^{*}<R^{*}.
    \label{eqn:FlowDim4}
 \end{equation}
\end{linenomath}
This fixes $H^{*} = 2V^{*}/\pi R^{*2}$ and hence we require that $\delta = 2V^{*}/\pi R^{*3}\ll1$.

Assuming $h^{*}>0$ for $0<r^{*}<R^{*}$ for $0<t^{*}<t_{f}^{*}$, where $t_{f}^{*}$ is the evaporation or dryout time of the drop \cite[][]{Deegan2000}, integrating (\ref{eqn:FlowDim1}a) from $r^{*} = 0$ to $r^{*} = R^{*}$ and applying the no-flux conditions (\ref{eqn:FlowDim2}b) and (\ref{eqn:FlowDim3}b), we obtain an expression representing global conservation of mass of the liquid phase, namely
\begin{linenomath}
 \begin{equation}
 \frac{\mbox{d}}{\mbox{d}t^{*}}\int_{0}^{R^{*}}r^{*}h^{*}(r^{*},t^{*})\,\mbox{d}r^{*} = - \int_{0}^{R^{*}}\frac{r^{*}E^{*}(r^{*})}{\rho^{*}}\,\mbox{d}r^{*};
 \end{equation}
\end{linenomath}
integrating and applying the initial condition (\ref{eqn:FlowDim4}) then gives
\begin{linenomath}
 \begin{equation}
 \int_{0}^{R^{*}}r^{*}h^{*}(r^{*},t^{*})\,\mbox{d}r^{*} = \int_{0}^{R^{*}} r^{*}h^{*}(r^{*},0)\,\mbox{d}r^{*} - \left(\int_{0}^{R^{*}}\frac{r^{*}E^{*}(r^{*})}{\rho^{*}}\,\mbox{d}r^{*}\right)t^{*},
 \end{equation}
\end{linenomath}
so that,
\begin{linenomath}
 \begin{equation}
 t_{f}^{*} = \frac{V^{*}}{2\pi}\left(\int_{0}^{R^{*}}\frac{r^{*}E^{*}(r^{*})}{\rho^{*}}\,\mbox{d}r^{*}\right)^{-1}.
 \label{eqn:EvaporationTimeDim}
 \end{equation}
\end{linenomath}
For a given evaporative flux, (\ref{eqn:FlowDim1})--(\ref{eqn:FlowDim4}) and (\ref{eqn:EvaporationTimeDim}) fully specify the flow problem for $h^{*}(r^{*},t^{*})$, $\bur^{*}(r^{*},t^{*})$, $p^{*}(r^{*},t^{*})$ and $t_{f}^{*}$. 

\subsubsection{Solute problem}

Since we are assuming that the dilute solute has no effect on the liquid flow, the solute concentration $\phi^{*}$ simply satisfies an advection-diffusion equation. In the limit in which $\delta\ll1$ and $\delta^{2}U^{*}R^{*}/D_{\phi}^{*}\ll1$ where $D_{\phi}^{*}$ is the solutal diffusion coefficient, it is straightforward to show that $\phi^{*}$ is independent of $z^{*}$ at leading order. Hence averaging the advection-diffusion equation across the droplet thickness yields
\begin{linenomath}
 \begin{equation}
  \frac{\partial}{\partial t^{*}}\left(h^{*}\phi^{*}\right) + \frac{1}{r^{*}}\frac{\partial}{\partial r^{*}}\left[r^{*}h^{*}\phi^{*}\bar{u}_{r}^{*} - D_{\phi}^{*}r^{*}h^{*}\frac{\partial\phi^{*}}{\partial r^{*}}\right] = 0
  \label{eqn:SolDim1}
 \end{equation}
\end{linenomath}
for $0<r^{*}<R^{*}$, $t^{*}>0$ \cite[][]{Wray2014, Pham2017}. There are two clear competing physical effects in (\ref{eqn:SolDim1}). The first is an advective flux of solute mass $\phi^{*}h^{*}\bar{u}_{r}^{*}$, where the evaporation-induced flow carries solute to the contact line. The solute concentration thus increases local to the contact line, in turn driving a competing diffusive flux $-D_{\phi}^{*}h^{*}\partial\phi^{*}/\partial r^{*}$ towards the droplet bulk. The relative importance of these effects is what drives the formation of the nascent coffee ring in our analysis. We have schematically illustrated these competing fluxes in figure \ref{fig:DropletConfigAxi}.

By symmetry, we have
\begin{linenomath}
 \begin{equation}
  \frac{\partial\phi^{*}}{\partial r^{*}} = 0 \quad \mbox{at} \quad r^{*} = 0, \; t^{*}>0,
  \label{eqn:SolDim2}
 \end{equation}
\end{linenomath}
while there can be no flux of particles through the contact line, so that
\begin{linenomath}
\begin{equation}
r^{*}h^{*}\phi^{*}\bar{u}_{r}^{*} - D_{\phi}^{*}r^{*}h^{*}\frac{\partial\phi^{*}}{\partial r^{*}} = 0 \quad \mbox{at} \quad r^{*} = R^{*}, \; t^{*}>0.
\label{eqn:SolDim3}
\end{equation}
\end{linenomath}
Finally, the initial solute distribution is taken to be uniform and given by
\begin{linenomath}
 \begin{equation}
  \phi^{*}(r^{*},0) = \phi_{\mathrm{init}}^{*} \quad \mbox{for} \quad 0<r^{*}<R^{*}.
  \label{eqn:SolDim4}
 \end{equation}
\end{linenomath}
For a given flow, (\ref{eqn:SolDim1})--(\ref{eqn:SolDim4}) completely specify the solute transport problem for $\phi^{*}(r^{*},t^{*})$.

We emphasize that, in the thin-droplet, dilute-solute limit, the solute problem decouples from the flow problem, so that we can solve (\ref{eqn:FlowDim1})--(\ref{eqn:FlowDim4}), (\ref{eqn:EvaporationTimeDim}) for $\bur^{*}$, $p^{*}$, $h^{*}$ and $t_{f}^{*}$, before solving for the solute concentration $\phi^{*}$ from (\ref{eqn:SolDim1})--(\ref{eqn:SolDim4}).

To close the problem, we require the velocity field $\bur^{*}$, which depends on the evaporative flux via (\ref{eqn:FlowDim1}). The evaporative flux (and hence the induced flow and structure) is dependent on the dominant evaporative process. However, we stress that, in our analysis, we are interested in the structure of the solute concentration profile near the contact line rather than determining which evaporative model is the most appropriate for a given problem. To that end, we shall consider two well-established evaporative models in this paper: a \textit{kinetic evaporation model} and a \textit{diffusive evaporation model}, as we shall now describe.

%%%%%% UP TO HERE %%%%%%

\subsubsection{Evaporation models}

In a kinetic evaporation model, it is often the case that either the surrounding gas consists entirely of the droplet vapour, or that diffusion away from the droplet surface happens sufficiently quickly that evaporation is limited by the liquid phase alone. As described in, for example, \citet{Murisic2011}, the evaporative flux in this regime is well approximated by the expression $ E^{*} = \mathcal{A}_{k}^{*} / (h^{*}+\mathcal{B}_{k}^{*})$, where $\mathcal{A}_{k}^{*}$ and $\mathcal{B}_{k}^{*}$ depend on the properties of the droplet and thermodynamic properties of the system. For a thin droplet, \cite{Murisic2011}, note that there are liquid/substrate systems (such as water/silicon) for which $\mathcal{B}_{k}^{*} \gg h^{*}$, so that $E^{*}\approx \mathcal{A}_{k}^{*}/\mathcal{B}_{k}^{*}$, a constant. For the purposes of this paper we therefore take the evaporative flux in this regime to be
\begin{linenomath}
 \begin{equation}
  E^{*}(r^{*}) = \mathcal{E}_{k}^{*}
  \label{eqn:KineticFluxDim}
 \end{equation}
\end{linenomath}
for $0<r^{*}<R^{*}$, where $\mathcal{E}_{k}^{*}$ is a constant. 

A constant evaporative flux may also be a reasonable approximation in other regimes for which kinetic evaporation is not the dominant effect. For example, \citet{Boulogne2016} consider water droplets evaporating on a glass substrate resting in a large hydrogel bath. Since the hydrogel dries at a similar speed to the water, the vapour concentration in the surrounding gas is greatly affected by the bath. For sufficiently large baths, the flux around the droplet is well-approximated by a constant flux, although \citet{Boulogne2016} note that convective effects may also be important in the air.

In a diffusive evaporation model, the dominant transport of the liquid vapour away from the droplet-air interface is diffusion, and for a wide range of problems, the vapour P\'{e}clet number is sufficiently large that this process is quasi-steady \cite[][]{Deegan2000,Hu2002}. The vapour concentration thus satisfies a mixed boundary value problem for Laplace's equation in the air. Since the droplet is thin, this problem is equivalent to solving for the potential outside a charged disk in classical electrostatics \cite[][]{Sneddon1966}, with the resulting evaporative flux given by
\begin{linenomath}
\begin{equation}
 E^{*}(r^{*}) = \frac{2R^{*}\mathcal{E}_{d}^{*}}{\pi}\frac{1}{\sqrt{R^{*2}-r^{*2}}}, \label{eqn:DiffusiveFluxDim}
\end{equation}
\end{linenomath}
for $0<r^{*}<R^{*}$, where again $\mathcal{E}_{d}^{*}$ depends on the properties of the system \cite[][]{Murisic2011}.

\subsection{Non-dimensionalization} \label{sec:NonDim}

The evaporative flux $E^{*}$ has a typical scale $\mathcal{E}^{*}$ that is given by $\mathcal{E}_{k}^{*}$ and $\mathcal{E}_{d}^{*}$ in the kinetic and diffusive evaporation models respectively. The induced radial velocity is $U^{*} = \mathcal{E}^{*}/\rho^{*}\delta$. Hence, we non-dimensionalize (\ref{eqn:FlowDim1})--(\ref{eqn:FlowDim4}), (\ref{eqn:EvaporationTimeDim}), (\ref{eqn:SolDim1})--(\ref{eqn:SolDim4}) and (\ref{eqn:KineticFluxDim})--(\ref{eqn:DiffusiveFluxDim}) by setting:
\begin{linenomath}\begin{equation}\begin{aligned}
 (r^{*}, & \, z^{*}) =  \; R^{*}( r,\eps z), \; \bur^{*} = 
\frac{\mathcal{E}^{*}}{\eps\rho^{*}}\bar{u}_{r}, \; t^{*}_{f} = \frac{\delta\rho^{*} R^{*}}{\mathcal{E}^{*}}t_{f}, \; t^{*} = \frac{\delta\rho^{*} R^{*}t_{f}}{\mathcal{E}^{*}}t & \\
  & h^{*} = \eps R^{*} h, \; p^{*} = p_{\mathrm{atm}}^{*} + \frac{\mu^{*}\mathcal{E}^{*}}{\rho^{*} R^{*}\eps^{3}} p, \; E^{*} = \mathcal{E}^{*} E, \; \phi^{*} = \phi_{\mathrm{init}}^{*}\phi. & 
  \label{eqn:Scalings}
\end{aligned}\end{equation}\end{linenomath}
We note that in (\ref{eqn:Scalings}), the dimensionless dryout time $t_{f}$ is given by
\begin{linenomath}
 \begin{numcases}
  {t_{f} = \frac{1}{4}\left(\int_{0}^{1} rE(r) \,\mbox{d}r\right)^{-1} =}
   \displaystyle{\frac{1}{2}} & \mbox{in the kinetic regime,} \label{eqn:EvaporationTimeKin}
\vspace{0.05in}\\
   \displaystyle{\frac{\pi}{8}} & \mbox{in the diffusive regime,}
     \label{eqn:EvaporationTimeDiff}
   \end{numcases}
\end{linenomath}
and we have rescaled time to fix its domain to be $0<t<1$, which will simplify substantially numerous expressions in the sequel.

Under the scalings (\ref{eqn:Scalings}), the flow problem (\ref{eqn:FlowDim1})--(\ref{eqn:FlowDim4}) becomes
\refstepcounter{equation} 
\begin{linenomath}
$$
 \frac{1}{t_{f}}\frac{\partial h}{\partial t} + \frac{1}{r}\frac{\partial}{\partial r}\left(rh\bar{u}_{r}\right) = -E, \quad  \bar{u}_{r} = -\frac{h^{2}}{3}\frac{\partial p}{\partial r}, \quad p = -\frac{1}{\Ca}\frac{1}{r}\frac{\partial}{\partial r}\left(r\frac{\partial h}{\partial r}\right)
 \eqno{(\theequation{\mathit{a},\mathit{b},\mathit{c}})}
 \label{eqn:FlowNonDim1}
$$
\end{linenomath}
for $0<r<1$, $t>0$, such that
\refstepcounter{equation} 
\begin{linenomath}
 $$
 \frac{\partial h}{\partial r} = rh\bar{u}_{r} = 0 \quad \mbox{at} \quad r = 0, \; t>0,
 \eqno{(\theequation{\mathit{a},\mathit{b}})}
 \label{eqn:FlowNonDim2}
 $$
\end{linenomath}
and
\refstepcounter{equation} 
\begin{linenomath}
 $$
 h = rh\bar{u}_{r} = 0 \quad \mbox{at} \quad r = 1, \; t>0,
 \eqno{(\theequation{\mathit{a},\mathit{b}})}
 \label{eqn:FlowNonDim3}
 $$
\end{linenomath}
along with the initial condition
\begin{linenomath}
\begin{equation}
 h(r,0) = 1-r^{2} \quad \mbox{for} \quad 0<r<1,
\end{equation}
\end{linenomath}
where,
\begin{linenomath}
 \begin{equation}
  \Ca = \frac{\mathcal{E}^{*}\mu^{*}}{\sigma^{*}\rho^{*}\eps^{4}}
  \label{eqn:Ca}
 \end{equation}
\end{linenomath}
is an enhanced droplet capillary number.

Similarly, the solute problem (\ref{eqn:SolDim1})--(\ref{eqn:SolDim4}) becomes
\begin{linenomath}
 \begin{equation}
  \frac{1}{t_{f}}\frac{\partial}{\partial t}\left(h\phi\right) + \frac{1}{r}\frac{\partial}{\partial r}\left[rh\phi\bar{u}_{r} - \frac{rh}{\Pe}\frac{\partial\phi}{\partial r}\right] = 0
  \label{eqn:SolNonDim1}
 \end{equation}
\end{linenomath}
for $0<r<1$, $t>0$, such that
\refstepcounter{equation}
\begin{linenomath}
 $$
  \frac{\partial\phi}{\partial r} = 0 \quad \mbox{at} \quad r = 0, \; t>0, \quad rh\phi\bar{u}_{r} - \frac{rh}{\Pe}\frac{\partial\phi}{\partial r} = 0 \quad \mbox{at} \quad r = 1, \; t>0,
   \eqno{(\theequation{\mathit{a},\mathit{b}})}
  \label{eqn:SolNonDim2}
 $$
\end{linenomath}
and initially
\begin{linenomath}
 \begin{equation}
  \phi(r,0) = 1 \quad \mbox{for} \quad 0<r<1.
  \label{eqn:SolNonDim3}
 \end{equation}
\end{linenomath}
Here, the solutal P\'{e}clet number, $\Pe$, is defined by
\begin{linenomath}
 \begin{equation}
  \Pe = \frac{\mathcal{E}^{*}R^{*}}{\rho^{*} D_{\phi}^{*}\eps}.
 \label{eqn:Pe}
 \end{equation}
\end{linenomath}

Finally, the dimensionless evaporative flux is given for $0<r<1$ by
\begin{linenomath}
 \begin{numcases}{E(r) =}
1 & \mbox{in the kinetic regime,} \label{eqn:KineticFlux} \vspace{0.05in}\\
          \displaystyle{\frac{2}{\pi}\frac{1}{\sqrt{1-r^{2}}}} & \mbox{in the diffusive regime.} \label{eqn:DiffusiveFlux}
 \end{numcases}
\end{linenomath}

\subsection{Small capillary number limit}

There are two dimensionless parameters in the problem: $\Ca$ and $\Pe$. We shall consider the regime in which $\Ca\ll1$, so that the fluid motion is dominated by surface tension. We shall then proceed to consider the large-$\Pe$ sublimit of this model, so that the importance of solute diffusion is confined to a region near the pinned contact line. Firstly, however, we will briefly illustrate the validity and applicability of these limits by considering several experimental studies from the literature. 

For a wide range of physical problems, the relevant limit for the flow model is that in which the droplet capillary number is small, so that surface tension dominates viscous forces in determining the droplet profile. To illustrate the pertinence of this assumption, consider a typical example from \citet{Hu2002}: a water droplet of volume and radius $V^{*} = 0.5 \mu$l and $R^{*} = 1 $mm respectively sits on a glass substrate and evaporates diffusively (cf. \textsection \ref{sec:Diffusive}) into the surrounding air. The typical flow velocities induced by the evaporation are reported to be on the order of $U^{*} = 1 \mu$ms$^{-1}$. Thus, $\delta \approx 0.3$ and  $\Ca \approx 4.3\times10^{-7}\ll1$. 

We see similar orders of magnitude for the droplet aspect ratio and the capillary number for other common liquids as well. \citet{Kajiya2008} study the coffee ring structure for evaporating droplets of anisole, for which $\mu^{*} = 1.03\times10^{-3}$Pa$\cdot$s and $\sigma = 3.5\times10^{-2}$Nm$^{-1}$. Again, the anisole droplet has volume and radius $V^{*} = 0.5 \mu$l and $R^{*} = 1 $mm respectively, while the authors report a drying time of $\approx 8$ minutes, which gives a fluid velocity $U^{*}\sim 0.8 \mu$ms$^{-1}$. Hence, for this example, we find that $\delta \approx 0.3$ and $\Ca \approx 7.47\times10^{-7}$.

As a final example, we consider the experimental and numerical analysis of evaporating ethanol droplets in \citet{Saenz2017}. The authors present experiments for a wide variety of droplet geometries, with the circular geometry of interest to the present study concerning droplets of radius $R^{*} \approx 2$mm and volume $V^{*} = 7 \mu$l. Although the authors do not explicitly estimate the induced flow velocity, they state that the final evaporation time of the droplets is $O(100)$s, which leads to $U^{*}\approx 10\mu$ms$^{-1}$. Thus, since ethanol has viscosity $\mu^{*} = 1.15\times10^{-3}$Pa$\cdot$s and surface tension $2.1\times10^{-2}$Nm$^{-1}$, we find that $\delta\approx0.6$ and the droplet capillary number is given by $\Ca\approx 2.49\times10^{-6}$.

Hence, as we see, for a wide variety of different liquids, we are very much entrenched in the small-$\Ca$ regime, so that in what follows, we shall assume that surface tension dominates the fluid motion.

For all of these examples, we have seen that $R^{*}\approx 1$mm and $U^{*}\approx 1-10 \mu$ms$^{-1}$. Hence, perhaps unsurprisingly, the size of the solutal P\'{e}clet number is most dependent on the size of the diffusion coefficient, $D^{*}_{\phi}$. If we return to the evaporating anisole droplets in \citet{Kajiya2008}, in which the solute is a fluorescent polystyrene for which the authors note $D_{\phi}^{*} \approx 2\times10^{-11}$m$^{2}$s$^{-1}$, we find that $\Pe \approx 40$.

More generally, if we assume that the solute particles are approximately spherical and hence that $D_{\phi}^{*}$ is well-modelled by the Stokes-Einstein equation, it is the size of the solute particles that is the dominant factor in determining the size of the solutal P\'{e}clet number (note that for all three fluids considered above, the viscosities were approximately $10^{-3}$Pa$\cdot$s). Thus, if we consider solute particle radii varying from $10 \mu$m (e.g. polystyrene microspheres, as in \citet{Deegan2000}) to $5\times10^{-4} \mu$m (e.g. a sugar molecule), the P\'{e}clet number varies from $\Pe\sim 3$ to $\Pe\sim5\times 10^{4}$. Thus, it is clear our assumption that $\Pe\gg1$ is applicable to a wide range of problems.

It is worth noting that, in the examples considered above, $\delta$ is not particularly small initially, so that we must be careful in the assumption of small-reduced P\'{e}clet number, $\eps^{2}\Pe$, that is made in writing down (\ref{eqn:SolDim1}). However, during evaporation the droplet necessarily becomes more slender, so that this assumption becomes more reasonable. 

The above estimates motivate us to pursue a small-$\Ca$, large-$\Pe$ solution of (\ref{eqn:FlowNonDim1})--(\ref{eqn:SolNonDim3}). Firstly, let us consider the flow in the droplet. As $\Ca\rightarrow0$, it is evident that\refstepcounter{equation}
\begin{linenomath}
\begin{eqnarray}
 h & \sim & \left(1-t\right)(1-r^{2}), \label{eqn:SmallCaSolutiona}\\
 p & \sim & \frac{4}{\Ca}(1-t), \label{eqn:SmallCaSolutionb}\\
 \bur & \sim & -\frac{1}{4t_{f}r}\frac{(1-r^{2})}{(1-t)} + \frac{1}{r(1-t)(1-r^{2})}\int_{r}^{1}sE(s)\,\mbox{d}s
 \label{eqn:SmallCaSolutionc}
\end{eqnarray}
\end{linenomath}
to leading order \cite[][]{Deegan2000}. Inspecting the final term in (\ref{eqn:SmallCaSolutionc}) reveals that, depending on the exact form of the evaporative flux, the depth-averaged radial velocity may be bounded as $r\rightarrow1$ (for (\ref{eqn:KineticFlux})) or singular as $r\rightarrow1$ (for (\ref{eqn:DiffusiveFlux})). The behaviour of $\bur$ as $r\rightarrow1$ will play an instrumental role in determining the structure of the diffusive boundary layer governed by (\ref{eqn:SolNonDim1})--(\ref{eqn:SolNonDim3}) close to the contact line, further justifying our consideration of these two possible behaviours in \textsection\textsection \ref{sec:Kinetic}--\ref{sec:Diffusive}.

\subsection{Formulation in terms of the solute mass}

Having determined the leading-order flow in the droplet, we turn to the solute problem. We assume that $\bur$ and $h$ are well-approximated by their leading-order forms in the small-$\Ca$ expansion and now concentrate on determining the asymptotic solution to (\ref{eqn:SolNonDim1})--(\ref{eqn:SolNonDim3}) in the limit in which $\varepsilon := 1/\Pe \rightarrow0^{+}$. Note, to be asymptotically consistent with our derivation of (\ref{eqn:SolNonDim1})--(\ref{eqn:SolNonDim3}), we are therefore considering the regime in which $\delta^{2}\ll\ve\ll1$. While the concentration, $\phi$, is important in determining when the dilute regime breaks down within the boundary layer (as we discuss in detail in \textsection \ref{sec:Jamming}), we find it is convenient to proceed with our asymptotic analysis by introducing the solute mass per unit area, $m = \phi h$, which satisfies
\begin{linenomath}
\begin{equation}
 \frac{1}{t_{f}}\frac{\partial m}{\partial t} + \frac{1}{r}\frac{\partial}{\partial r}\left[r\left(\bur + \frac{\ve}{h}\frac{\partial h}{\partial r}\right)m - \ve r\frac{\partial m}{\partial r}\right] = 0 \quad \mbox{for} \quad 0<r<1, \; t>0,
 \label{eqn:MassEqn}
\end{equation} 
\end{linenomath}
subject to
\refstepcounter{equation}
\begin{linenomath}
$$
 \frac{\partial m}{\partial r} = 0 \; \mbox{at} \; r = 0, \; t>0, \quad 
 r\left(\bur + \frac{\ve}{h}\frac{\partial h}{\partial r}\right)m - \ve r\frac{\partial m}{\partial r} = 0 \quad \mbox{at} \quad r = 1, \; t>0
 \eqno{(\theequation{\mathit{a},\mathit{b}})}
 \label{eqn:MassBC}
$$
\end{linenomath}
and 
\begin{linenomath}
\begin{equation}
 m(r,0) = h(r,0) = 1-r^{2} \quad \mbox{for} \quad 0<r<1. \label{eqn:MassIC}
\end{equation}
\end{linenomath}
Finally, since it will be useful in what follows, we note that global conservation of solute dictates that
\begin{linenomath}
 \begin{equation}
  \int_{0}^{1} rm(r,t)\,\mbox{d}r = \int_{0}^{1} rm(r,0)\,\mbox{d}r = \frac{1}{4}.
  \label{eqn:ConsOfSolute}
 \end{equation}
\end{linenomath}

We also note that, in addition to being mathematically convenient, using the solute mass has an advantage over $\phi$ because it is related to the absorbance of the deposit via the Beer-Lambert law \cite[][]{Swinehart1962}; it may therefore be easier to compare predictions of $m$ directly to experimental data.

As mentioned previously, the asymptotic structure is sensitive to the behaviour of the radial velocity, which is given by (\ref{eqn:SmallCaSolutionc}), close to the contact line. For a kinetic evaporative flux, the velocity is bounded at the contact line, which we consider in \textsection \ref{sec:Kinetic}. On the other hand, for a diffusive evaporative flux, the velocity is singular at the contact line, and we consider this case in \textsection \ref{sec:Diffusive}. Our goal is to formulate a composite expansion for the solute mass valid everywhere in the droplet to leading order and to use the asymptotic results to establish thereby the dynamics of the nascent coffee ring for each flux law. We will compare these predictions to numerical simulations of the full system (\ref{eqn:MassEqn})--(\ref{eqn:MassIC}) (aided by a further reformulation that is motivated, described and assessed in Appendix \ref{sec:NumericalMethods}).

%%%%%%%%%%%%%%%%%%%%%%%%%%%%%%%%%%%%%%%%%%%%%%%%%%%%%%%%%%%%%%%
%%%%%%%%%%%%%%%%%%%%%%%%% KINETIC FLUX %%%%%%%%%%%%%%%%%%%%%%%%
%%%%%%%%%%%%%%%%%%%%%%%%%%%%%%%%%%%%%%%%%%%%%%%%%%%%%%%%%%%%%%%

\section{Boundary layer structure for kinetic evaporation} \label{sec:Kinetic} 

% In a kinetic evaporation model, it is often the case that either the surrounding gas entirely consists of the droplet vapour, or that diffusion away from the droplet surface happens sufficiently quickly that evaporation is limited by the liquid phase alone. While there are several different forms such an evaporative flux can take \cite[][]{Burelbach1988, Sultan2005, Murisic2011}, in the limit in which the droplet is thin, these are all well-approximated by a constant evaporative flux, so that in our dimensionless problem,
% \begin{linenomath}
%  \begin{equation}
%   E \equiv 1.
%  \end{equation}
% \end{linenomath}
Substituting the kinetic evaporative flux (\ref{eqn:KineticFlux}) and the dryout time (\ref{eqn:EvaporationTimeKin}) into (\ref{eqn:SmallCaSolutionc}) yields
\begin{linenomath}
\begin{equation}
  \bur = \frac{r}{2(1-t)}. \label{eqn:hurpsKin}
\end{equation}
\end{linenomath}
We may then proceed to seek an asymptotic solution of (\ref{eqn:MassEqn}), (\ref{eqn:MassBC}b) and (\ref{eqn:MassIC}) as $\ve\rightarrow0$. We note at the outset that it is straightforward to show that the outer solution we find satisfies the symmetry condition (\ref{eqn:MassBC}a), so we do not need to introduce a boundary layer at $r = 0$.

\subsection{Outer region} \label{sec:KinOuter}

In the outer region, we na\"{i}vely expand $m = m_{0} + \ve m_{1} + O(\ve^{2})$ as $\ve\rightarrow0$, where, as we shall shortly see, we must proceed to $O(\ve)$ in the outer region in order to be able to construct a composite mass solution that vanishes at the contact line. To leading order in $\ve$, (\ref{eqn:MassEqn}) and (\ref{eqn:MassIC}) become
\begin{linenomath}
 \begin{equation}
  \frac{\partial m_{0}}{\partial t} + \frac{1}{r}\frac{\partial}{\partial r}\left(\frac{r^{2}}{4(1-t)}m_{0}\right) = 0 \quad \mbox{for} \quad 0<r<1, \; t>0,  \label{eqn:LOOKin}
 \end{equation}
\end{linenomath}
with $m_{0}(r,0) = 1-r^{2}$ for $0<r<1$. Hence, as expected, the leading-order solute mass is simply advected to the contact line by the radial flow. We can solve (\ref{eqn:LOOKin}) using the method of characteristics, finding
\begin{linenomath}
 \begin{equation}
  m_{0}(r,t) = a(t)\left(1-r^{2}a(t)\right),
  \label{eqn:M0Kin}
 \end{equation}
\end{linenomath}
where we have introduced the function $a(t) = \sqrt{1-t}$. We note that (\ref{eqn:M0Kin}) can be used to determine, to leading order in $\ve$, the total amount of solute swept into $r = 1$ by time $t$, denoted by $\mathcal{M}(t)$, namely
\begin{linenomath}
 \begin{equation}
  \mathcal{M}(t) = t_{f}\int_{0}^{t} \bur(1^{-},\tau)m_{0}(1^{-},\tau)\,\mbox{d}\tau = \frac{1}{2}\left(1-\sqrt{1-t}-\frac{t}{2}\right),
  \label{eqn:MassFluxInKin}
 \end{equation}
\end{linenomath}
which was previously reported for a kinetic evaporative flux by \citet{FreedBrown2015} and \citet{Boulogne2016}.

At $O(\ve)$, we have
\begin{linenomath}
 \begin{equation}
  \frac{\partial m_{1}}{\partial t} + \frac{1}{r}\frac{\partial}{\partial r}\left(\frac{r^{2}}{4(1-t)}m_{1}\right) = \frac{2a(t)(1-a(t))}{(1-r^{2})^{2}} \quad \mbox{for} \quad 0<r<1, \; t>0,
 \end{equation}
\end{linenomath}
where $m_{1}(r,0) = 0$ for $0<r<1$. This can be solved in a similar manner, yielding
\begin{linenomath}
\begin{eqnarray}
 m_{1}(r,t) & = & \frac{1}{r}\frac{\partial}{\partial r}\left[r^{2}a(t)\left((1-r^{2}a(t))t - \frac{2}{3}\left(1-a(t)^{3}\right) - \right. \right. \nonumber \\
 & & \left. \left. 2a(t)r^{2}\left(1-r^{2}a(t)\right)\left(a(t) - 1 + a(t)r^{2}\log\left(\frac{a(t)(1-r^{2})}{1-a(t)r^{2}}\right)\right)\right)\right]. \label{eqn:M1Kin}
\end{eqnarray}
\end{linenomath}

Note that, as $r\rightarrow1$, $m_{0} \rightarrow a(t)(1-a(t))$ and $m_{1} = O(1/(1-r))$, both of which are physically unreasonable, as the mass should vanish at the contact line because the droplet thickness vanishes there. This is a clear indication of the need to consider the behaviour close to $r = 1$, where the effects of solute diffusion become relevant.

\subsection{Inner region} \label{sec:KineticInner}

We scale into the inner region by setting
\begin{linenomath}
 \begin{equation}
  r = 1 - \ve R, \; m = \ve^{-1}M
 \end{equation}
\end{linenomath}
in (\ref{eqn:MassEqn})--(\ref{eqn:MassIC}), where the scaling for the mass has been determined from the global conservation of mass condition, (\ref{eqn:ConsOfSolute}). Then, to account for the logarithmic terms in the local expansion of (\ref{eqn:M1Kin}) at the contact line, we seek an asymptotic series of the form
\begin{linenomath}
 \begin{equation}
  M = M_{0} + (\ve\log\ve) M_{1} + \ve M_{2} +  O(\ve^{2}\log\ve)
 \end{equation}
\end{linenomath}
as $\ve\rightarrow0$. The leading-order inner problem is given by
\begin{linenomath}
 \begin{equation}
  \frac{\partial}{\partial R}\left[\left(\alpha(t)-\frac{1}{R}\right) M_{0} + \frac{\partial M_{0}}{\partial R}\right]=0 \quad \mbox{for} \quad R>0, \; t>0,
  \label{eqn:LOInnerKin1}
 \end{equation}
\end{linenomath}
such that
\begin{linenomath}
 \begin{equation}
  \left(\alpha(t)-\frac{1}{R}\right)M_{0} +\frac{\partial M_{0}}{\partial R} = 0 \quad \mbox{at} \quad R = 0, \; t>0,
  \label{eqn:LOInnerKin2}
 \end{equation}
\end{linenomath}
where we have introduced the function $\alpha(t) = 1/(2(1-t))$. Equations (\ref{eqn:LOInnerKin1})--(\ref{eqn:LOInnerKin2}) are readily solved, yielding
\begin{linenomath}
 \begin{equation}
  M_{0} = A(t) R\exp{\left(-\alpha(t) R\right)},
  \label{eqn:tM0Kin}
 \end{equation}
\end{linenomath}
where $A(t)$ is a function of time that we shall determine shortly. The functional form of $M_{0}$ --- which may be viewed as the probability density function for a gamma distribution --- drives the characteristic nascent coffee ring profile in the boundary layer. We shall discuss this further in \textsection \ref{sec:CoffeeRingProperties}.

At $O(\ve\log\ve)$, the inner problem is the same as that at leading order, so that
\begin{linenomath}
 \begin{equation}
  M_{1} = B(t) R\exp{\left(-\alpha(t) R\right)},
  \label{eqn:tM1Kin}
 \end{equation}
\end{linenomath}
where $B(t)$ is to be determined. 

At $O(\ve)$, we find that
\begin{linenomath} 
 \begin{equation}
 \frac{\partial}{\partial R}\left[\left(\alpha(t)-\frac{1}{R}\right)M_{2} + \frac{\partial M_{2}}{\partial R}\right] = 2\frac{\partial M_{0}}{\partial t} - \frac{\partial}{\partial R}\left[\left(\frac{3}{2}-2\alpha(t) R\right) M_{0} - R\frac{\partial M_{0}}{\partial R}\right] \label{eqn:Order2InnerKin}
\end{equation}
\end{linenomath}
in $R>0$, $t>0$, such that
\begin{linenomath}
 \begin{equation}
  \left(\alpha(t)-\frac{1}{R}\right)M_{2} + \frac{\partial M_{2}}{\partial R} + \left(\frac{3}{2}-2\alpha(t) R\right) M_{0} - R\frac{\partial M_{0}}{\partial R} = 0 \quad \mbox{at} \quad R = 0, \; t>0.
 \end{equation}
\end{linenomath} 
The solution to this problem is given by
\begin{linenomath}
\begin{equation}
 M_{2} = \left[C_{1}(t)R + C_{2}(t)R^{2} + C_{3}(t)R^{3} + C_{4}(t)R\left(\mbox{Ei}(\alpha(t) R) - \log(\alpha(t) R)\right) \right]\exp\left(-\alpha(t) R\right)
 \label{eqn:tM2Kin}
\end{equation}
\end{linenomath}
where $\mbox{Ei}(x)$ is the exponential integral, $C_{1}(t)$ is to be determined and 
\begin{linenomath}
\begin{equation}
 C_{2}(t) = \frac{15}{2}A(t) - \frac{2}{\alpha(t)}\dot{A}(t), \; C_{3}(t) = \frac{5\alpha(t)}{2} A(t), \; C_{4}(t) = \frac{2}{\alpha(t)^{2}}\dot{A}(t) - \frac{8}{\alpha(t)}A(t).
 \label{eqn:CisKin}
\end{equation}
\end{linenomath}

\subsection{Determining the unknown coefficients}

%%%% Truncate this to a paragraph and say we could theoretically get A, B, C_{1} by matching and have enough for A, we would need to go to even higher order to get B, C_{1} which unwieldy -> use conserv. of solute.
It remains to determine the unknown coefficients $A(t)$, $B(t)$ and $C_{1}(t)$ in (\ref{eqn:tM0Kin}), (\ref{eqn:tM1Kin}) and (\ref{eqn:tM2Kin}) respectively. In principle, we can obtain these by matching between the inner and outer solutions, although this is notably challenging as (\ref{eqn:tM0Kin}) and (\ref{eqn:tM1Kin}) decay exponentially in the far-field, so we must proceed to higher order to complete the matching. We already have enough information to match for $A(t)$ in this manner, but we would need to go to even higher order in the inner region to match for $B(t)$ and $C_{1}(t)$, which becomes algebraically taxing. To avoid this work, we can instead appeal to global conservation of solute, (\ref{eqn:ConsOfSolute}). The details of this approach are given in Appendix \ref{sec:ConSolAlg}, and show that
\begin{linenomath}
 \begin{equation}
  A(t) = \frac{1}{4(1-t)^{2}}\left[\frac{1}{2}-\frac{\sqrt{1-t}}{2}-\frac{t}{4}\right], \; B(t) = \frac{1-\sqrt{1-t}}{2\sqrt{1-t}} 
  \label{eqn:beta_and_A}
 \end{equation}
\end{linenomath}
and
\begin{linenomath}
\begin{eqnarray}
 C_{1}(t) & = & -\alpha(t)^{2}a(t)\left[(1-a(t))t - \frac{2}{3}\left(1-a(t)^{3}\right) -\right. \nonumber \\
 & & \left. 2a(t)\left(1-a(t)\right)\left(a(t) - 1 + a(t)\log\left(\frac{2a(t)}{1-a(t)}\right)\right)\right] - \nonumber \\
 & &  \frac{2(C_{2}(t)-A(t))}{\alpha(t)} - \frac{6C_{3}(t)}{\alpha(t)^{2}} - 2C_{4}(t)(\gamma-1) - C_{4}(t)\log\alpha(t).
 \label{eqn:C1}
\end{eqnarray}
\end{linenomath}
Note that the term in square brackets in $A(t)$ is simply the total solute mass advected into the boundary layer from the outer region at leading-order, (\ref{eqn:MassFluxInKin}).

% Our first course of action is to match between the inner and outer solutions. This is most readily accomplished by introducing an intermediate variable,
% \begin{linenomath}
%  \begin{equation}
%   r = 1 - \ve^{k}\sukr = 1 - \ve R \; \mbox{where} \; 0<k<1.
%   \label{eqn:Intermediate}
%  \end{equation}
% \end{linenomath}
% Substituting this intermediate scaling into the first two terms of the outer expansion given by (\ref{eqn:M0Kin}) and (\ref{eqn:M1Kin}) gives
% \begin{linenomath}
%  \begin{equation}
%   M_{outer} = \frac{a}{4}(1-a) + \ve^{1-k}\frac{a^{3}}{8\sukr}(1-a) + O(\ve)
%  \end{equation}
% \end{linenomath}
% as $\ve\rightarrow0$. Similarly, substituting (\ref{eqn:Intermediate}) into the first three terms of the inner solution (\ref{eqn:tM0Kin}), (\ref{eqn:tM1Kin}) and (\ref{eqn:tM2Kin}), all contributions from the first two terms are negligible, while the third gives
% \begin{linenomath}
%  \begin{equation}
%   M_{inner} = \frac{C_{4}}{\alpha} + \ve^{1-k}\frac{C_{4}}{\alpha^{2}} +  O(\ve^{\beta_{1}+2}).
%  \end{equation}
% \end{linenomath}

% Since they are coefficients of exponentially decaying terms, in order to obtain expressions for $B(t)$ and $C_{1}(t)$ by matching, we would need to proceed to even higher order in the inner region, which becomes algebraically taxing. We can circumvent this difficulty by appealing to global conservation of solute, (\ref{eqn:ConsOfSolute}). 

\subsection{Composite expansion}

Having determined the unknown coefficients in the inner and outer expansions, we are left to form a composite mass profile. A natural first attempt would be to simply write down
\begin{linenomath}
 \begin{equation}
  m_{\mathrm{comp}} = m_{0}(r,t) + \frac{1}{\ve} M_{0}\left(\frac{1-r}{\ve},t\right),
  \label{eqn:MCompKinBad}
 \end{equation}
\end{linenomath}
where $m_{0}$ and $M_{0}$ are given by (\ref{eqn:M0Kin}) and (\ref{eqn:tM0Kin}) respectively. As we shall see, while (\ref{eqn:MCompKinBad}) does a fine job in approximating the mass profile in the droplet bulk, close to the contact line, there is an $O(1)$ error in the mass profile caused by the fact that $m_{0}$ is finite as $r\rightarrow1$.

As forecast by our consideration of the higher-order outer and inner problems, to remedy this we can write down a second composite expansion that is valid up to $O(1)$ for all $r$. This entails using $m_{0}, M_{0}, M_{1}$ and $M_{2}$. To form the composite, we turn to Van Dyke's matching rule \cite[][]{VanDyke1964}, which requires a knowledge of the common contribution from the outer and inner solutions.  We introduce the intermediate variable $r = 1 - \ve^{k}\bar{r} = 1-\ve R$, where $0<k<1$. Substituting this scaling into the inner and outer solutions and expanding as $\ve\rightarrow0$, we see that
\refstepcounter{equation}
\begin{linenomath}
 $$
  m_{\mathrm{inner}} = \frac{C_{4}(t)}{\alpha(t)^{2}} + o(1) = a(t)(1-a(t)) + o(1), \quad
  m_{\mathrm{outer}} = a(t)(1-a(t)) + o(1),
  \eqno{(\theequation{\mathit{a},\mathit{b}})}
  \label{eqn:DummyMatchingKin}
 $$
\end{linenomath}
where we have exploited the known value of $A(t)$ in the inner expansion to evaluate the expression in (\ref{eqn:DummyMatchingKin}a). 

Therefore, the appropriate composite expansion valid to $O(1)$ for all $r$ is
\begin{linenomath}
\begin{eqnarray}
 m_{\mathrm{comp}} & = & m_{0}(r,t) +  \frac{1}{\ve} M_{0}\left(\frac{1-r}{\ve},t\right) + \nonumber \\
 & &  (\log\ve)M_{1}\left(\frac{1-r}{\ve},t\right) + M_{2}\left(\frac{1-r}{\ve},t\right) - a(t)(1-a(t)),
 \label{eqn:MCompKin}
\end{eqnarray}
\end{linenomath}
where $M_{1}$ and $M_{2}$ are given by (\ref{eqn:tM1Kin}) and (\ref{eqn:tM2Kin}) respectively and we recall $a(t) = \sqrt{1-t}$. We note that, although we do not need the precise form of $m_{1}$ to write down this composite solution, we did need to ascertain its behaviour as $r\rightarrow1$ to perform the correct asymptotic expansion in the inner region, which is why we considered the $O(\ve)$-outer problem in \textsection \ref{sec:KinOuter}. Moreover, the logarithmic singularity in $m_{1}$ given by (\ref{eqn:M1Kin}) makes it challenging to write down a higher-order composite expansion that is valid everywhere.

\subsection{Profiles and properties of the coffee ring} \label{sec:CoffeeRingProperties}

Now we have determined two composite solutions for the solute mass, we are able to demonstrate the formation and evolution of the nascent coffee ring effect as solute is advected from the outer region into the diffusive boundary layer, which grows in a characteristic gamma distribution-type profile.

In the following, an important quantity is the modified time-dependent P\'{e}clet number (henceforth the modified P\'{e}clet number): 
\begin{linenomath}
 \begin{equation}
  \Pe_{t} = \frac{\Pe}{1-t},
  \label{eqn:LocalPeclet}
 \end{equation}
\end{linenomath}
which is the ratio of advective to diffusive transport modified to take into account the time dependence of the evaporation-induced velocity, the latter scaling with the reciprocal of the time to dryout, i.e. $(1-t)^{-1}$. Since we shall see that at leading order the thickness and height of the diffusive boundary layer scale with functions of the modified P\'{e}clet number, it captures in a single quantity the scaling behaviour of the nascent coffee ring with respect to both the P\'{e}clet number, $\Pe$,  and the time remaining until dryout.

Of particular note is that our asymptotic analysis suggests that the form of the nascent coffee ring is dominated by the leading-order-inner solution (\ref{eqn:tM0Kin}) for $\Pe_{t}\gg1$. In particular, we expect the nascent coffee ring to tend to the similarity form given by
\begin{linenomath}
 \begin{equation}
  \frac{M_{0}(R,t)}{\mathcal{M}(t)\Pe_{t}} = \frac{R}{4}\mbox{e}^{-R/2} = f\left(R;2,\frac{1}{2}\right), \; R = \Pe_{t}(1-r), 
  \label{eqn:KineticSimilarity}
 \end{equation}
\end{linenomath}
where $\mathcal{M}(t)$ is given by (\ref{eqn:MassFluxInKin}) and $f(x,k,l) = l^{k}x^{k-1}\mbox{e}^{-l x}/\Gamma(k)$ is the probability density function of a gamma distribution.

To validate our asymptotic predictions, we have also solved (\ref{eqn:MassEqn})--(\ref{eqn:MassIC}) numerically. This solution is hindered by the boundary layer thickness and height scaling with $1/\Pe_{t}$ and $\Pe_{t}$ as $\Pe_{t}\rightarrow\infty$; we have ensured that this boundary layer is resolved by discretizing in a suitable manner, as described in Appendix \ref{sec:NumericalMethods}. 

\begin{figure}
\begin{subfigure}
\centering \scalebox{0.375}{\epsfig{file=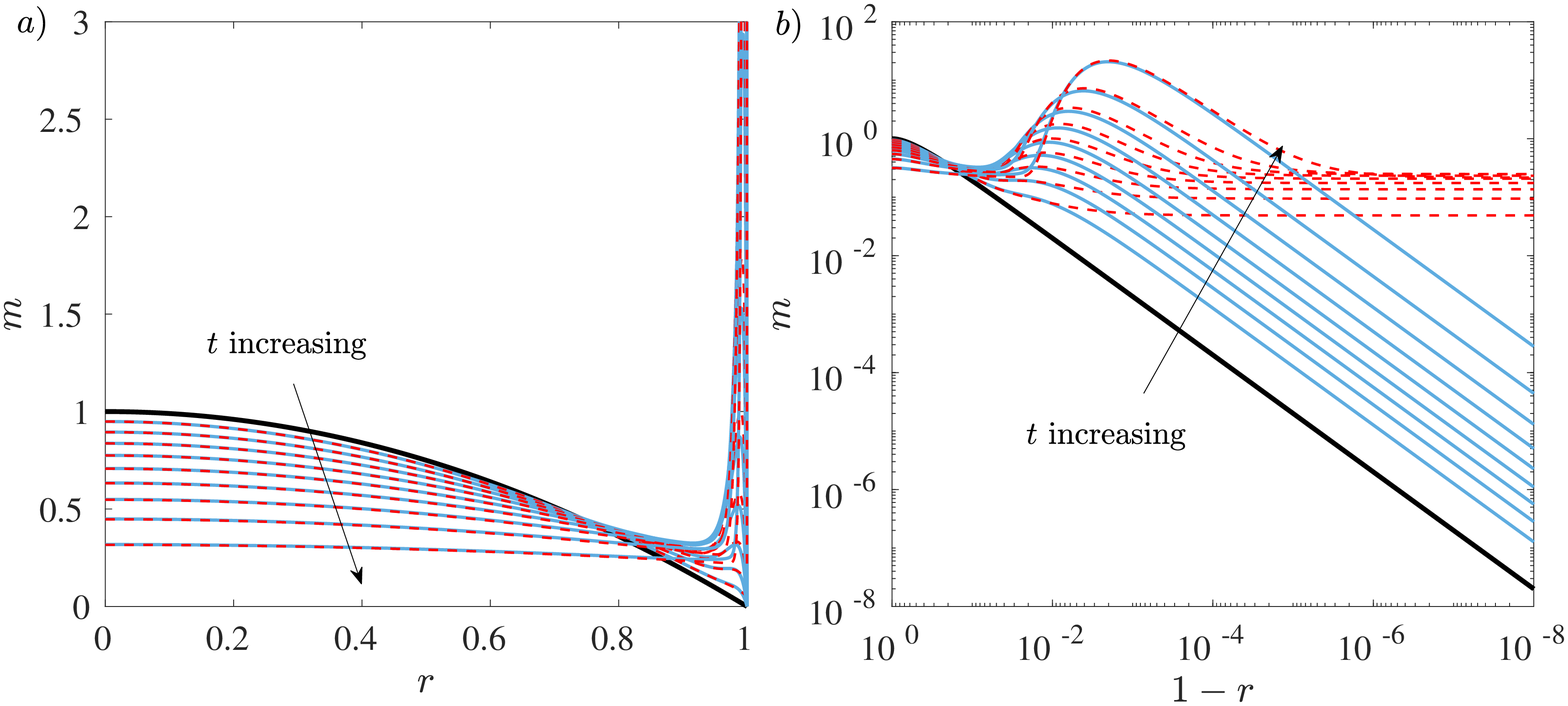}}
\captionsetup{labelformat=empty}
\end{subfigure}
\begin{subfigure}
\centering \scalebox{0.375}{\epsfig{file=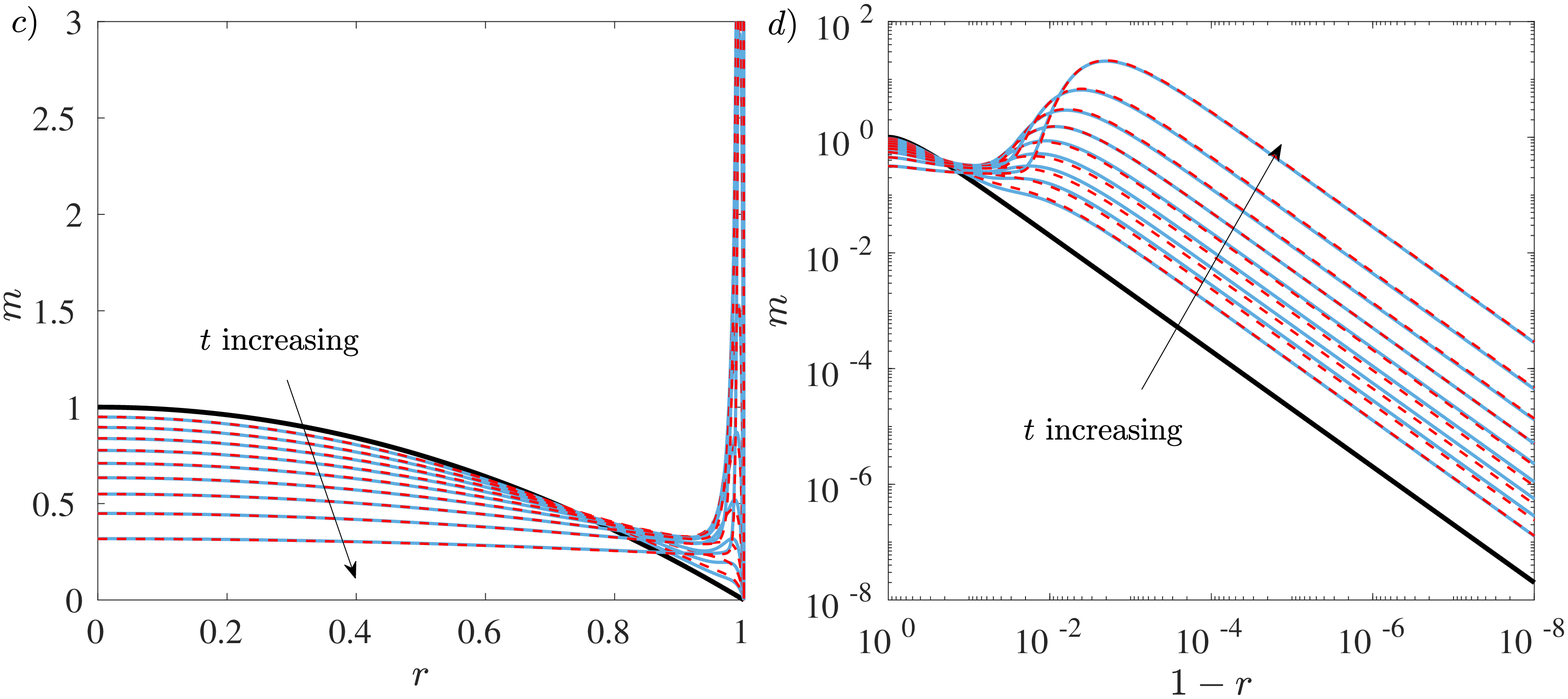}}
\end{subfigure}
\caption{Profiles of the solute mass as the droplet evaporates under a kinetic flux, $E = 1$, with $\Pe = 100$. In each figure, the bold, black curve represents the initial mass profile (which is identical to the droplet profile since $\phi(r,0) = 1$). Also shown are plots at time intervals of 0.1 up to $t = 0.9$ in which solid, blue curves represent the numerical results and the dashed, red curves show the composite mass profiles predicted by (\ref{eqn:MCompKinBad}) ($a, b$) and (\ref{eqn:MCompKin}) ($c, d$) for large-$\Pe$. Figures $b, d$ display a doubly-logarithmic plot of the mass profile near the contact line, where we can clearly see the formation of the concentration spike that becomes the nascent coffee ring as $t$ increases.}
\label{fig:ProfileComparisons} 
\end{figure}

We plot both the numerical and asymptotic predictions of the solute mass profile in figure \ref{fig:ProfileComparisons} for $\Pe = 100$. In the top row, we compare the composite mass solution given by (\ref{eqn:MCompKinBad}) to the numerical solution, while in the bottom row we show the comparison for the composite mass profile given by (\ref{eqn:MCompKin}). In each plot, we see the evolution of the solute mass from the initial profile (bold, black curve). As the droplet evaporates, the geometry of the droplet induces a radial flow towards the contact line, which advects solute outwards. In the boundary layer, diffusion leads to the characteristic coffee-ring spike, which is clearly seen in the insets to each plot. While it is clear that both composite profiles do an excellent job of capturing the solute in the droplet bulk, there is a clear deviation of the lower-order composite (\ref{eqn:MCompKinBad}) from the expected solution in the boundary layer, leading to $O(1)$ errors. The higher-order inner solutions accounted for in (\ref{eqn:MCompKin}) rectify this deficiency, and we see excellent agreement between the asymptotics and the numerics, particularly as $t$ increases. We note in particular that, at 90\% of the dryout time (the final curve in each plot), the peak mass in the boundary layer is approximately $22$ times the initial peak at the centre of the drop (for which $m_{\mathrm{max}} = 1$). 

\begin{figure}
\centering \scalebox{0.4}{\epsfig{file=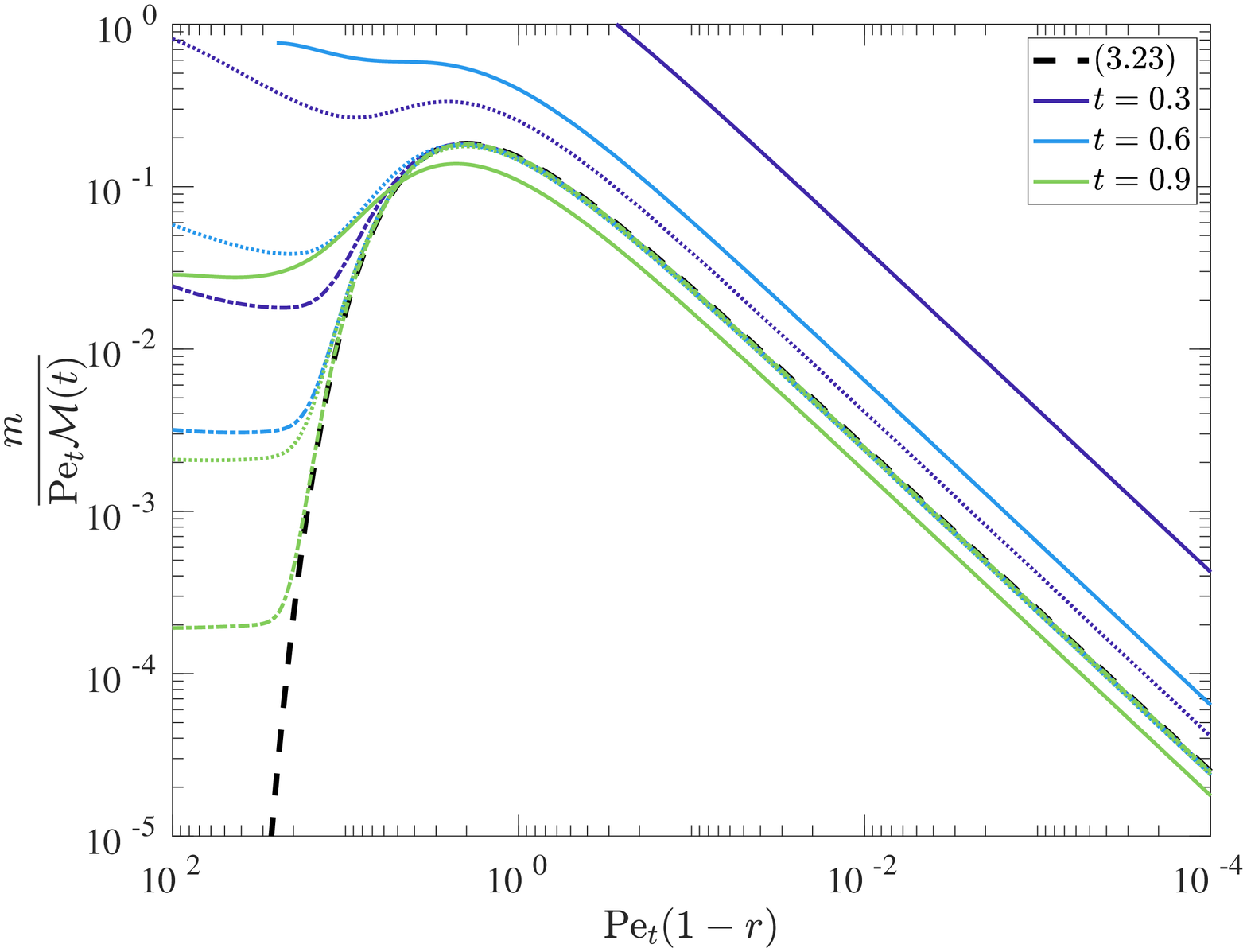}}
\caption{The scaled solute mass $m/\mathcal{M}(t)\Pe_{t}$ against the radial similarity coordinate $\Pe_{t}(1-r)$ as a function of time. The dashed, black curve represents the scaled leading-order-inner solute mass profile (\ref{eqn:KineticSimilarity}). The coloured curves are results from the numerical simulations at $t = 0.3$ (dark purple), $t = 0.6$ (blue) and $t = 0.9$ (light green) for $\Pe = 10$ (solid), $\Pe = 100$ (dots) and $\Pe = 1000$ (dash-dots). As $t\rightarrow1$ and $\Pe\rightarrow\infty$, the results collapse onto the similarity form given by (\ref{eqn:KineticSimilarity}), as expected.}
\label{fig:CollapseKinetic} 
\end{figure}

We depict in figure \ref{fig:CollapseKinetic} the anticipated collapse of the nascent coffee ring onto the similarity form (\ref{eqn:KineticSimilarity}). For a wide range of $\Pe$, we see clear evidence of this similarity form emerging after an initial transient, demonstrating the universality of the gamma distribution profile (\ref{eqn:KineticSimilarity}) in the diffusive boundary layer.

\begin{figure}
\centering \scalebox{0.31}{\epsfig{file=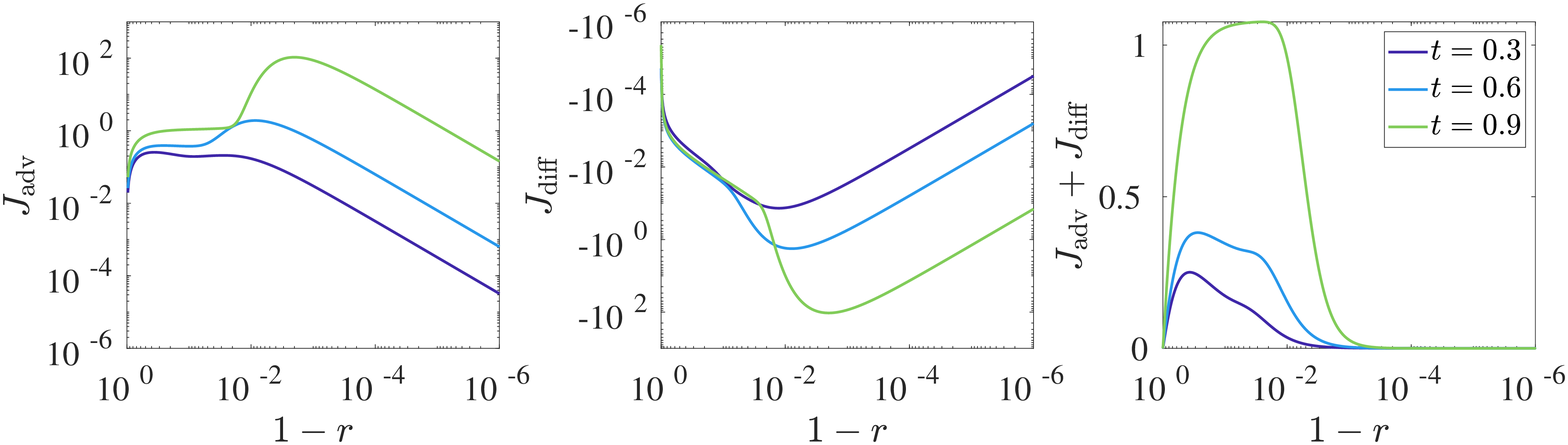}}
\caption{Transient profiles of the advective particle flux $J_{\mathrm{adv}}$ (left), the diffusive particle flux $J_{\mathrm{diff}}$ (centre) and the overall mass flux (right) for a droplet evaporating in the kinetic evaporative flux regime with $\Pe = 100$ at times $t = 0.3$ (dark purple), $t = 0.6$ (blue) and $t = 0.9$ (light green). These profiles have been calculated from the composite solution (\ref{eqn:MCompKin}).}
\label{fig:FluxesKinetic} 
\end{figure}

Since we are now armed with sufficient evidence of the validity of our asymptotic results, we can use them to investigate various aspects of the solutal flow dynamics and the development of the nascent coffee ring. In particular, this analysis was motivated by establishing the competition between the advective mass flux $J_{\mathrm{adv}} = \bur h\phi$ and the diffusive mass flux $J_{\mathrm{diff}} = -(h/\Pe)\partial\phi/\partial r$ in different regions of the droplet. We display both of these leading-order fluxes and the resulting leading-order mass flux at various times for $\Pe = 100$ in figure \ref{fig:FluxesKinetic}. As is clearly seen, the advective flux carries solute particles towards the contact line, while the diffusive flux acts to take particles away from the high concentration at the contact line. The magnitude of the maximum advective and diffusive fluxes is comparable at each instant, which is expected from the asymptotic structure of the model: in the boundary layer, the dominant balance is between these fluxes at the expense of the time derivative term in (\ref{eqn:MassEqn}). In each of the profiles shown in figure \ref{fig:FluxesKinetic}, it is clear that the advective flux is larger, so that the resultant movement of solute is towards the pinned contact line, which leads to the development of the coffee ring. At each time instant, there is a slow increase of the overall mass flux towards the contact line, which rapidly falls as we enter the boundary region where the diffusive flux is important. It should be noted that, as we can see in figure \ref{fig:FluxesKinetic}, the peak mass flux moves towards the contact line as $t\rightarrow1$; this is indicative of the majority of mass having been advected radially out from the droplet centre as it dries. Moreover, it is clear that, while the overall mass flux towards the contact line generally increases in magnitude as $t$ increases, the \emph{relative} mass flux has fallen: at time $t = 0.9$, the maximum advective and diffusive fluxes are roughly three orders of magnitude larger than those at $t  = 0.3$, while the maximum overall mass flux has increased by just a factor of $4$.

\begin{figure}
\centering \scalebox{0.4}{\epsfig{file=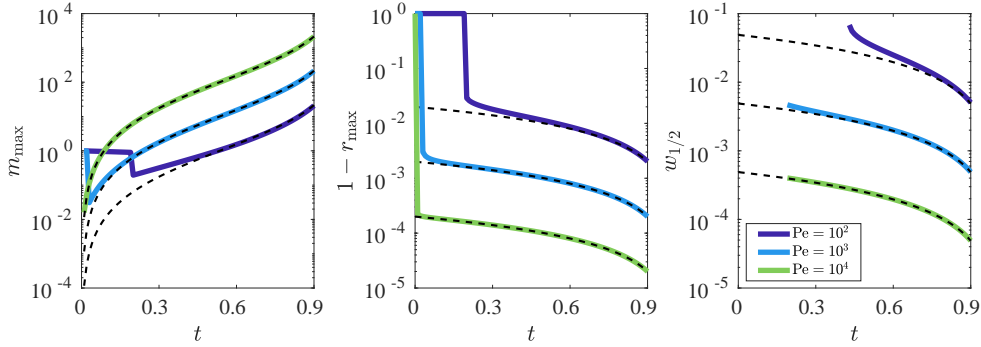}}
\caption{The maximum height of the solute mass profile $m_{\mathrm{max}}$ (left), its location $1-r_{\mathrm{max}}$ (centre) and the profile full-width at half-maximum $w_{1/2}$ (right) as a function of time in the kinetic evaporative flux regime. In this figure, we consider the numerical results for three different P\'{e}clet numbers, $\Pe = 10^{2}$ (dark purple), $\Pe = 10^{3}$ (blue), $\Pe = 10^{4}$ (light green). For each case, the dashed, black curve corresponds to the asymptotic predictions (\ref{eqn:hmax_and_rmax}) and (\ref{eqn:whalf}). Note that $w_{1/2}$ is only well-defined for $t \gtrsim 0.2$ ($t \gtrsim 0.5$ for $\Pe = 100$); see main text for details.}
\label{fig:CRProps} 
\end{figure}

We conclude by using our asymptotic results to predict characteristics of the nascent coffee ring that may be of use in applications and, in particular, measurable in an experimental setting. As it forms, the maximum height, $m_{\mathrm{max}}$, of the solute profile and the radial position of the maximum, $r_{\mathrm{max}}$, can be well approximated by maximizing the leading-order-inner solution, (\ref{eqn:tM0Kin}), yielding
\begin{linenomath}
 \begin{equation}
  m_{\mathrm{max}}(t) = \frac{\mathcal{M}(t)\Pe_{t}}{2\mbox{e}}, \; r_{\mathrm{max}}(t) = 1 - \frac{2}{\Pe_{t}},
  \label{eqn:hmax_and_rmax}
 \end{equation}
\end{linenomath}
where $\mathcal{M}(t)$ is given by (\ref{eqn:MassFluxInKin}). In particular, we note that, as the droplet evaporates, the peak moves closer to the contact line and the maximum solute profile height diverges as $t\rightarrow1$. This is illustrated for three different P\'{e}clet numbers in figure \ref{fig:CRProps}. The figure also demonstrates that the asymptotics provide a good approximation of the numerical results at an earlier time for larger P\'{e}clet numbers, as expected. We note that the sharp changes in the figure correspond to the maximum in the solute mass profile moving from the droplet centre to the nascent coffee ring in the boundary layer. This happens at earlier times for larger P\'{e}clet numbers.

Finally, we consider a measure of the size of the coffee ring by considering the width of the profile at half its peak height --- the full-width at half-maximum --- which we denote by $w_{1/2}$. Asymptotically, this is given by
\begin{linenomath}
\begin{equation}
 w_{1/2}(t) = \frac{2}{\Pe_{t}}\left[W_{0}\left(\frac{-1}{2\mbox{e}}\right) - W_{-1}\left(\frac{-1}{2\mbox{e}}\right)\right],
 \label{eqn:whalf}
\end{equation}
\end{linenomath}
where $W_{0}(x)$ and $W_{-1}(x)$ are the Lambert-W functions (i.e solutions to $w\mbox{e}^{w} = x$); \cite[][]{Olver2010}. Clearly, the width of the nascent coffee ring shrinks as the drop evaporates, with the majority of the mass confined to a sharp, narrow peak --- note that (\ref{eqn:tM0Kin}) tends to a delta function, $\delta(R)$, as $t\rightarrow1$. We plot the asymptotic prediction (\ref{eqn:whalf}) in figure \ref{fig:CRProps} alongside the evolution of $w_{1/2}$ extracted from the numerical results. We note that results are only shown for $t$ such that this definition makes sense: for small times, the minimum in the solute mass profile between the coffee ring and the droplet bulk lies above half the maximum height, making $w_{1/2}$ ill-defined. As the P\'{e}clet number increases, $w_{1/2}$ is well-defined for a longer time period. As with the comparisons for $m_{\mathrm{max}}$ and $r_{\mathrm{max}}$, we see that the leading-order asymptotic prediction holds over a longer time interval as $\Pe$ increases, while even for the smallest P\'{e}clet number depicted, the asymptotic prediction captures the behaviour of the half-width $w_{1/2}$ well as $t\rightarrow1$.

%%%%%%%%%%%%%%%%%%%%%%%%%%%%%%%%%%%%%%%%%%%%%%%%%%%%%%%%%%%%%%%
%%%%%%%%%%%%%%%%%%%%%%%% DIFFUSIVE FLUX %%%%%%%%%%%%%%%%%%%%%%%
%%%%%%%%%%%%%%%%%%%%%%%%%%%%%%%%%%%%%%%%%%%%%%%%%%%%%%%%%%%%%%%

\section{Boundary layer structure for diffusive evaporation} \label{sec:Diffusive}

We now move on to consider the equivalent boundary layer structure and nascent coffee-ring properties under a diffusive evaporative flux, (\ref{eqn:DiffusiveFlux}). Notably, this flux is singular at the contact line, so that evaporation is strongest there, enhancing the liquid flow towards the contact line that is required to replace the lost fluid \cite[][]{Deegan1997}.

With the evaporative flux (\ref{eqn:DiffusiveFlux}) and the dryout time given by (\ref{eqn:EvaporationTimeDiff}), the depth-averaged radial velocity $\bur$ can be determined from (\ref{eqn:SmallCaSolutionc}), giving
\begin{linenomath}
 \begin{equation}
 \bur = \frac{2}{\pi r(1-t)}\left(\frac{1}{\sqrt{1-r^{2}}} - (1-r^{2})\right).
  \label{eqn:hurpsDiff}
  \end{equation}
\end{linenomath}
As in the kinetic regime, we now seek an asymptotic solution of (\ref{eqn:MassEqn}), (\ref{eqn:MassBC}b) and (\ref{eqn:MassIC}) as $\ve\rightarrow0$; note that, again the symmetry condition (\ref{eqn:MassBC}a) is automatically satisfied at leading order so a boundary layer is not needed at $r = 0$.

\subsection{Outer region}

In the outer region, we expand $m = m_{0} + O(\ve)$ as $\ve\rightarrow0$ in (\ref{eqn:MassEqn})--(\ref{eqn:MassIC}). Then, to leading order in $\ve$, we obtain the advection equation:
\begin{linenomath}
 \begin{equation}
  \frac{\partial m_{0}}{\partial t} + \frac{1}{r}\frac{\partial}{\partial r}\left[\frac{1}{4(1-t)}\left(\frac{1}{\sqrt{1-r^{2}}} - (1-r^{2})\right)m_{0}\right] = 0 \quad \mbox{for} \quad 0<r<1, \; t>0,
  \label{eqn:LOODiff}
 \end{equation}
\end{linenomath}
with $m_{0}(r,0) = (1-r^{2})$ for $0<r<1$.

Equation (\ref{eqn:LOODiff}) can again be solved by the method of characteristics: we find that
\begin{linenomath}
 \begin{equation}
  m_{0} = \sqrt{1-r^{2}}(1-t)^{3/4}\left[1-(1-t)^{3/4}(1-(1-r^{2})^{3/2})\right]^{1/3}.
  \label{eqn:M0Diff}
 \end{equation}
\end{linenomath}
In particular, we note that $m_{0}$ vanishes as $r\rightarrow1$, in contrast to the kinetic regime for which $m_{0}$ was bounded, but finite at the contact line. We will show that it is for this reason that we do not need to proceed to higher order here: the leading-order outer solution is sufficient to construct a composite mass profile that is accurate enough for our purposes. We also note that we can use (\ref{eqn:M0Diff}) to recover (to leading-order in $\ve$) the total mass of solute advected into $r = 1$ by the radial outward flow as a function of time, viz:
\begin{linenomath}
 \begin{equation}
  \mathcal{M}(t) = t_{f}\int_{0}^{t} \bur(1^{-},\tau)m_{0}(1^{-},\tau)\,\mbox{d}\tau = \frac{1}{4}\left(1-(1-t)^{3/4}\right)^{4/3},
  \label{eqn:MassFluxInDiff}
 \end{equation}
\end{linenomath}
a result previously reported in \citet{Deegan2000} and \citet{Popov2003}.

\subsection{Inner region} \label{sec:DiffusiveInner}

In the diffusive regime, we scale into the contact line region by setting
\begin{linenomath}
 \begin{equation}
  r = 1 - \ve^{2}R, \; m = \ve^{-2}M,
  \label{eqn:DiffusiveScalings}
 \end{equation}
\end{linenomath}
where, as previously, the scaling for $m$ can be determined from the conservation of solute condition, (\ref{eqn:ConsOfSolute}). We can see immediately that the diffusive boundary layer is an order of magnitude thinner, while the solute mass profile is an order of magnitude larger for this evaporation model than in the kinetic evaporation model (in which the width and height are of $O(\ve)$ and $O(\ve^{-1})$ respectively). This is consistent with the experimental results of \citet{Kajiya2008}, in which droplets allowed to evaporate naturally (i.e. diffusively) produced thinner coffee rings than those constrained to evaporate in a box (i.e. closer to the kinetic regime).

Upon substituting the scalings (\ref{eqn:DiffusiveScalings}) into (\ref{eqn:MassEqn})--(\ref{eqn:MassIC}) and expanding the mass in an asymptotic series of the form $M = M_{0} + O(\ve)$ as $\ve\rightarrow0$, we find to leading order that
\begin{linenomath}
 \begin{equation}
  \frac{\partial}{\partial R}\left[\left(\frac{1}{4\sqrt{2}(1-t)\sqrt{R}} - \frac{\pi}{8R}\right)M_{0} + \frac{\pi}{8}\frac{\partial M_{0}}{\partial R}\right] = 0 \quad \mbox{for} \quad R>0, \; t>0,
 \end{equation}
\end{linenomath}
such that
\begin{linenomath}
 \begin{equation}
  \left(\frac{1}{4\sqrt{2}(1-t)\sqrt{R}} - \frac{\pi}{8R}\right) M_{0} + \frac{\pi}{8}\frac{\partial M_{0}}{\partial R} = 0 \quad \mbox{at} \quad R=0, \; t>0.
 \end{equation}
\end{linenomath}
The leading-order-inner solute mass profile is therefore given by
\begin{linenomath}
 \begin{equation}
  M_{0} = F(t)R\mbox{exp}\left(-\frac{2\sqrt{2}}{\pi(1-t)}\sqrt{R}\right),
  \label{eqn:tM0Diff}
 \end{equation}
\end{linenomath}
where $F(t)$ is an arbitrary function of $t$.

\subsection{Determining the unknown coefficient $F(t)$}

Since the leading-order-inner solution is exponentially small as $R\rightarrow\infty$, we again appeal to conservation of solute to find an expression for $F(t)$. Although the evaporative flux has changed, conservation of solute (\ref{eqn:ConsOfSolute}) must still hold. We can then determine $F(t)$ in a similar manner to the kinetic regime, see Appendix \ref{sec:ConSolAlg}; we find that
\begin{linenomath}
\begin{equation}
 F(t) = \frac{16}{3\pi^{4}}\frac{1}{(1-t)^{4}}\left[\frac{1}{4}\left(1-(1-t)^{3/4}\right)^{4/3}\right].
 \label{eqn:beta2_and_F}
\end{equation}
\end{linenomath}
We have written this in a slightly unusual manner in order to highlight that, as with $A(t)$ in the kinetic regime, the term in square brackets is the total solute mass advected into the contact line region up to time $t$, (\ref{eqn:MassFluxInDiff}).

\subsection{Composite expansion}

We can write the leading-order composite mass solution for the diffusive evaporative flux model as
\begin{linenomath}
 \begin{equation}
  m_{comp} = m_{0}(r,t) + \frac{1}{\ve^{2}}M_{0}\left(\frac{1-r}{\ve^{2}},t\right),
  \label{eqn:MCompDiff}
 \end{equation}
\end{linenomath}
where $m_{0}$ and $M_{0}$ are given by (\ref{eqn:M0Diff}) and (\ref{eqn:tM0Diff}) respectively. To obtain a higher-order composite approximation for the solute mass in this regime requires tackling the second-order-outer problem, which will typically need to be done numerically. However, as we shall see, the leading-order composite expansion (\ref{eqn:MCompDiff}) is in excellent agreement with our numerical simulations.

\subsection{Profiles and properties of the coffee ring} \label{sec:CoffeeRingPropertiesDiff}

With a diffusive evaporative model, the stronger evaporative flux at the contact line leads to a sharper, thinner solute profile in the boundary layer. In particular, as the modified P\'{e}clet number, $\Pe_{t}$, increases, we expect the nascent coffee ring to approach the similarity form given by
\begin{linenomath}
 \begin{equation}
  \frac{M_{0}(R,t)}{\mathcal{M}(t)\Pe_{t}^{2}} = \frac{16R}{3\pi^{4}}\mbox{e}^{-2\sqrt{2}\sqrt{R}/\pi} = \frac{\sqrt{2}}{3\pi}f\left(\sqrt{R};3,\frac{2\sqrt{2}}{\pi}\right), \; R = \Pe_{t}^{2}(1-r),
  \label{eqn:DiffusiveSimilarity}
 \end{equation}
\end{linenomath}
where $\mathcal{M}(t)$ is given by (\ref{eqn:MassFluxInDiff}) and we again see the emergence of a gamma distribution probability density function, $f$, although with an increased shape parameter ($3$ rather than $2$ in the kinetic case).

\begin{figure}
\centering \scalebox{0.375}{\epsfig{file=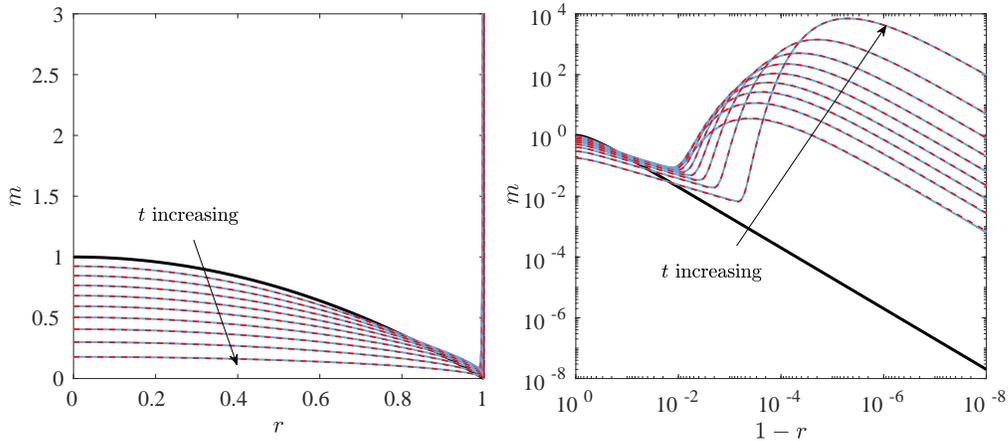}}
\caption{Profiles of the solute mass as the droplet evaporates under a diffusive flux given by (\ref{eqn:KineticFlux}) with $\Pe = 100$. In each figure, the bold, black curve represents the initial droplet solute profile (which mimics the initial droplet profile). Plots at time intervals of $0.1$ up to $t = 0.9$ show the results of numerical simulations (solid, blue curves) and the composite mass profile (dashed, red curves) predicted by our large-$\Pe$ asymptotics (\ref{eqn:MCompDiff}). The right-hand figure is a doubly-logarithmic plot of the mass profile near the contact line, where we can clearly see the formation of the nascent coffee ring as $t$ increases.}
\label{fig:ProfileComparisonsDiff} 
\end{figure}

To check the veracity of our asymptotic predictions, we again compare them to numerical simulations of the full system (\ref{eqn:MassEqn})--(\ref{eqn:MassIC}). The nature of the extremely thin boundary layer with diffusive evaporation --- recall, an order of magnitude thinner than the kinetic regime --- means more care has to be taken in the range of P\'{e}clet numbers considered to balance numerical convergence with remaining in the large-$\Pe$ asymptotic regime. In figure \ref{fig:ProfileComparisonsDiff}, we consider both numerical solutions (solid curves) and the asymptotic predictions (dashed curves) of the solute mass profiles as $t$ increases up to 90\% of the drying time for $\Pe = 100$. Again, we can clearly see the formation of the peak in the solute profile that should ultimately become the coffee ring as the drop evaporates. We note that the stronger evaporative flux in this regime induces a much sharper increase in the coffee ring height: in the inset to figure \ref{fig:ProfileComparisonsDiff}, we see that by $t = 0.9$, the solute mass is $\approx 7000$, which is two orders of magnitude larger than the height of the mass profile in the kinetic regime at the same stage of the dynamics. Nevertheless, we again see excellent agreement between the asymptotic predictions and the results of our numerical simulations.

\begin{figure}
\centering \scalebox{0.4}{\epsfig{file=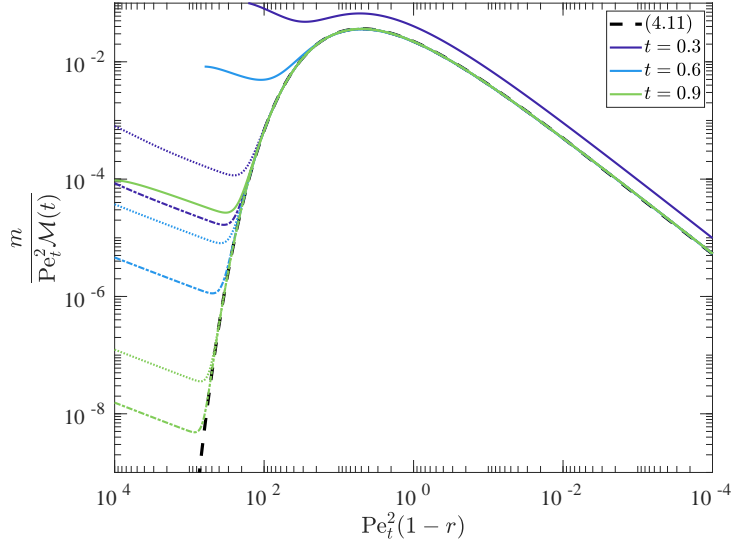}}
\caption{The scaled solute mass $m/\mathcal{M}(t)\Pe_{t}^{2}$ against the radial similarity coordinate $\Pe_{t}^{2}(1-r)$ as a function of time. The dashed, black curve represents the scaled leading-order-inner solute mass profile (\ref{eqn:DiffusiveSimilarity}). The coloured curves are results from the numerical simulations at $t = 0.3$ (dark purple), $t = 0.6$ (blue) and $t = 0.9$ (light green) for $\Pe = 10$ (solid), $\Pe = 100$ (dots) and $\Pe = 1000$ (dash-dots). We see the expected collapse to the similarity form (\ref{eqn:DiffusiveSimilarity}) as $\Pe_{t}$ increases.}
\label{fig:CollapseDiff} 
\end{figure}

In figure \ref{fig:CollapseDiff}, we demonstrate the collapse of the nascent coffee ring to (\ref{eqn:DiffusiveSimilarity}) as $\Pe_{t}$ increases. The convergence to the similarity profile is very rapid in the diffusive regime and gives us confidence, alongside the excellent comparisons of the composite mass profile in figure \ref{fig:ProfileComparisonsDiff}, in using our model to predict features of the nascent coffee ring.

\begin{figure}
\centering \scalebox{0.4}{\epsfig{file=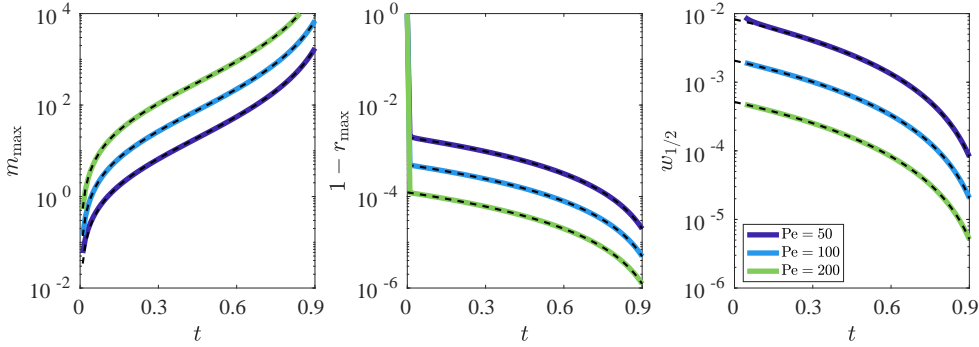}}
\caption{The maximum height of the solute mass profile $m_{\mathrm{max}}$ (left), its location $1-r_{\mathrm{max}}$ (centre) and the full-width at half-maximum $w_{1/2}$ (right) as a function of time in the diffusive regime. In this figure we consider three different P\'{e}clet numbers, $\Pe = 50$ (dark purple), $\Pe = 100$ (blue), $\Pe = 200$ (light green). For each case, the dashed, black curve corresponds to the asymptotic predictions (\ref{eqn:hmax_and_rmax_Diff}) and (\ref{eqn:whalfDiff}).}
\label{fig:CRPropsDiff} 
\end{figure}

The maximum height, $m_{\mathrm{max}}$, and its radial position, $r_{\mathrm{max}}$, can again be well approximated by considering the leading-order-inner solution $M_{0}$, yielding
\begin{linenomath}
 \begin{equation}
  m_{\mathrm{max}} = \frac{8\mathcal{M}(t)\Pe_{t}^{2}}{3\pi^{2}\mbox{e}^{2}}, \; r_{\mathrm{max}} = 1 - \frac{\pi^{2}}{2\Pe_{t}^{2}}.
  \label{eqn:hmax_and_rmax_Diff}
 \end{equation}
\end{linenomath}
As with the kinetic regime (\ref{eqn:hmax_and_rmax}), $m_{\mathrm{max}}$ diverges and $r_{\mathrm{max}}$ approaches the contact line as $t\rightarrow1$, although for diffusive evaporation, these effects are much more pronounced. This can clearly be seen in figure \ref{fig:CRPropsDiff}, where we plot the numerical and asymptotic predictions for $\Pe = 50, 100$ and $200$. In comparison to, in particular, the $\Pe = 100$ curve in figure \ref{fig:CRProps}, we see that the concentration peak forms much sooner in the diffusive regime than in the kinetic regime due to the enhanced effects of evaporation. We can also see that, even compared to the $\Pe = 1000$ case in figure \ref{fig:CRProps}, the concentration peak is much closer to the contact line in this regime, driven by the stronger velocity profile.

The width of the concentration peak at half of its maximum height in the diffusive regime can be found analytically to be
\begin{linenomath}
 \begin{equation}
  w_{1/2} = \frac{\pi^{2}}{2\Pe_{t}^{2}}\left[W_{-1}\left(\frac{-1}{\sqrt{2}\mbox{e}}\right)^{2} - W_{0}\left(\frac{-1}{\sqrt{2}\mbox{e}}\right)^{2}\right],
  \label{eqn:whalfDiff}
 \end{equation}
\end{linenomath}
which, consistent with the behaviours of $m_{\mathrm{max}}$ and $r_{\mathrm{max}}$, shrinks more rapidly than its equivalent in the kinetic regime (\ref{eqn:whalf}) as the droplet evaporates. We plot (\ref{eqn:whalfDiff}) alongside the corresponding numerical results in figure \ref{fig:CRPropsDiff}. Commensurate with the rapidity with which the nascent coffee ring forms in this regime, we are able to capture $w_{1/2}$ much sooner than in the kinetic regime, and we see that even for P\'{e}clet numbers as small as $50$, there is excellent agreement between the asymptotic prediction (\ref{eqn:whalfDiff}) and the numerical results.

Finally, we note that it is more difficult to display the competing fluxes in the diffusive evaporative regime than it was for the kinetic evaporative regime (cf. figure \ref{fig:FluxesKinetic}). This is primarily due to the fact that, while the composite asymptotic expansion (\ref{eqn:MCompDiff}) is asymptotically consistent for the solute mass, $m$, it loses this consistency if we attempt to differentiate it, which is necessary when computing $J_{\mathrm{diff}} = -(h/\Pe)\partial\phi/\partial r$. The inconsistency arises from the fact that, although the leading-order-outer solute mass (\ref{eqn:M0Diff}) is square-root bounded at the contact line, $\partial\phi/\partial r \sim (1-r)^{-3/2}$ as $r\rightarrow1$. Hence, $J_{\mathrm{diff}}$ is inverse square-root singular at the contact line. In order to address this singularity, we would need to proceed to higher order in both the inner and outer regions of our asymptotic analysis, which is a challenging procedure and beyond the scope of the present paper. However, as we shall now discuss, there are even more significant challenges facing the diffusive evaporative flux regime.

%%%%%%%%%%%%%%%%%%%%%%%%%%%%%%%%%%%%%%%%%%%%%%%%%%%%%%%%%%%%%%%
%%%%%%%%%%%%%%%%%%%%%%%%%%%% JAMMING %%%%%%%%%%%%%%%%%%%%%%%%%%
%%%%%%%%%%%%%%%%%%%%%%%%%%%%%%%%%%%%%%%%%%%%%%%%%%%%%%%%%%%%%%%

\section{Breakdown of the dilute approximation}
\label{sec:Jamming}

One result of the rapid formation of a thin, relatively-concentrated profile is that the concentration of solute is likely to rapidly reach the limits of the dilute approximation employed here. A variety of finite concentration effects may enter including concentration-dependent diffusivity or suspension viscosity, ultimately leading to solute jamming. We do not consider such effects here, but rather seek to use our analysis to understand when the dilute approximation is likely to break down. We suppose that the characteristic packing fraction at which the dilute approximation ceases to be valid is given by $\phi^{*} = \phi_{c}^{*}$ and we refer to this as the limiting concentration. The largest value of the (rescaled) solute concentration, $\phi_{\mathrm{max}}(t)$, is at the contact line, and we use the large-$\Pe$ asymptotics to approximate $\phi_{\mathrm{max}}(t)$ and compare it to the rescaled limiting concentration, $\phi_{c}^{*}/\phi_{\mathrm{init}}^{*}$. 

For the kinetic regime, we may use the composite mass profile (\ref{eqn:MCompKin}) to show that
\begin{linenomath}
 \begin{equation}
  \phi_{\mathrm{max}}(t) \sim \frac{\Pe^{2}}{2(1-t)}\left[A(t) + \frac{\log{1/\Pe}}{\Pe}B(t) + \frac{1}{\Pe}(C_{1}(t) + \gamma C_{4}(t))\right] + 1.
  \label{eqn:JammingKin}
 \end{equation}
\end{linenomath}

Matters are less straightforward for the diffusive regime: as alluded to at the end of \textsection \ref{sec:Diffusive}, since $m_{0}$ is only square-root bounded as $r\rightarrow1$, $m_{\mathrm{comp}}/h$ is unbounded as we approach the contact line. Instead, we simply use the leading-order-inner solute mass (\ref{eqn:tM0Diff}), yielding
\begin{linenomath}
 \begin{equation}
  \phi_{\mathrm{max}}(t) \sim \frac{2\Pe^{4}}{3\pi^{4}(1-t)^{5}}\left[1-(1-t)^{3/4}\right]^{4/3}.
  \label{eqn:JammingDiff}
 \end{equation}
\end{linenomath}
A drawback of having only the leading-order-inner term in this approximation is that as $t\rightarrow0$, we have $\phi_{\mathrm{max}}(0^{+}) \rightarrow0$, rather than $\phi_{\mathrm{max}}(0^{+}) = 1$, as demanded by the initial condition (\ref{eqn:SolNonDim3}). This deficiency arises because we have not proceeded to high enough order in the inner expansion to pick up order unity contributions to $\phi$. Nonetheless, we anticipate there to be a non-uniformity in our asymptotic expansion in the inner region as $t\rightarrow0^{+}$ since, at sufficiently small times, the $\partial M/\partial t$ term must enter the inner problem at leading order in order to satisfy the local form of the initial condition. Since this early-time deficiency is absent from our composite expansions for the solute mass and does not affect our asymptotic predictions for $t = O(1)$, we do not analyse it further here — except to emphasize the caveat concerning the deficiency of the approximation in the diffusive regime (\ref{eqn:JammingDiff}) as $t\rightarrow0^{+}$.

\begin{figure}
\centering \scalebox{0.45}{\epsfig{file=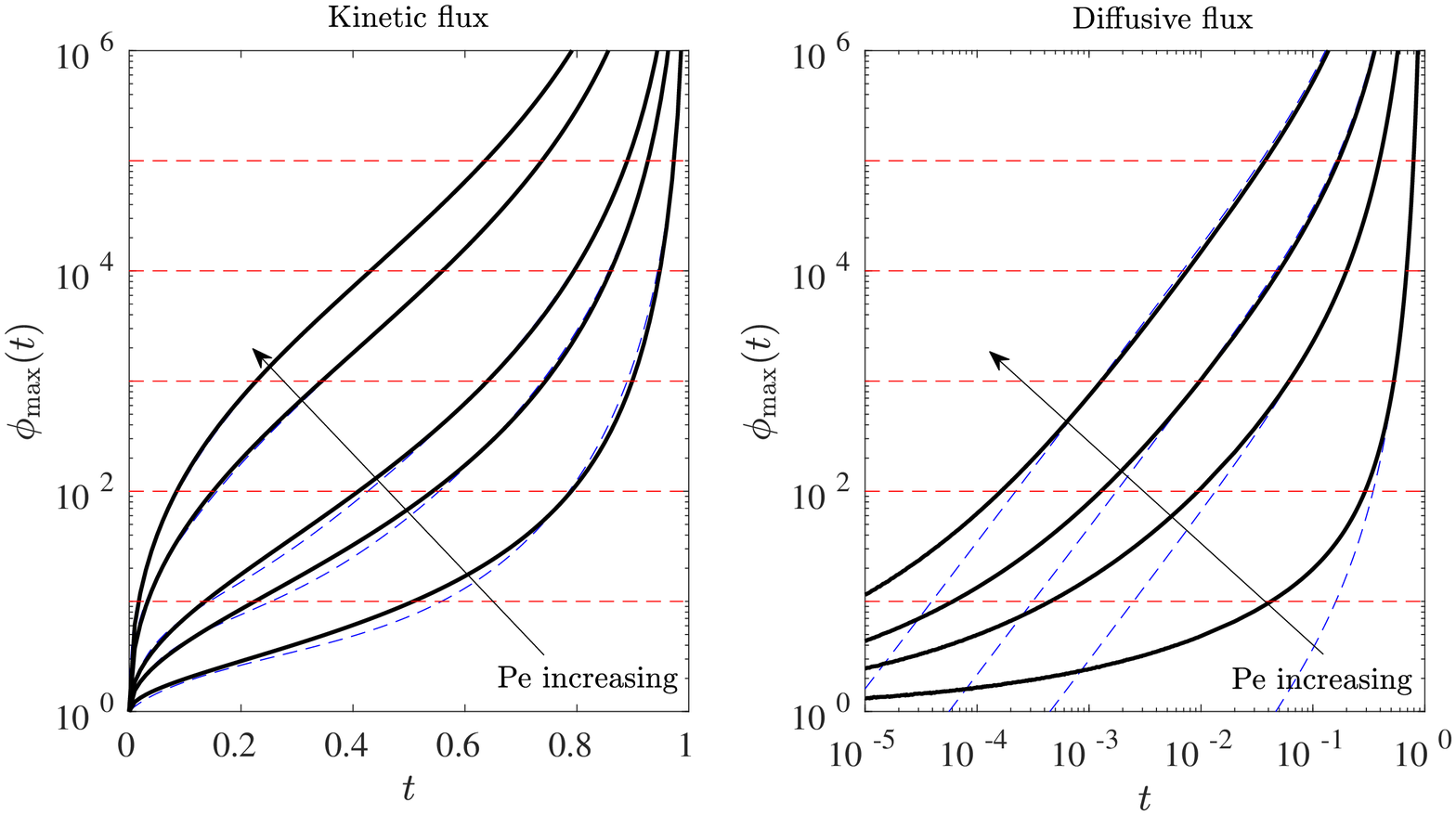}}
\caption{The evolution of the maximum solute concentration (that at the contact line) under the kinetic (left) and diffusive (right) evaporation models. We show the evolution of $\phi_{\mathrm{max}}(t)$ for $\Pe = 10, 50, 100, 500$ and $1000$ in the kinetic regime and for $\Pe = 10, 50, 100$ and $200$ in the diffusive regime. The dashed lines represent the asymptotic predictions given by (\ref{eqn:JammingKin}) and (\ref{eqn:JammingDiff}), while the solid lines depict the results of our numerical simulations. Our dilute analysis is expected to hold while $\phi_{\mathrm{max}}(t) < \phi_{c}^{*}/\phi_{\mathrm{init}}^{*}$ --- as illustrated by the dashed lines corresponding to $\phi_{c}^{*}/\phi^{*}_{\mathrm{init}} = 10^{2}-10^{5}$. Note that the diffusive case is plotted on a doubly-logarithmic scale to highlight the extremely rapid growth of $\phi_{\mathrm{max}}(t)$ predicted by (\ref{eqn:JammingDiff}).}
\label{fig:Jamming} 
\end{figure}

This caveat notwithstanding, in figure \ref{fig:Jamming} we plot these maximum concentrations as functions of time for different P\'{e}clet numbers varying from $10-1000$. Since the critical value $\phi_{c}^{*}/\phi_{\mathrm{init}}^{*}$ varies depending both on the solute under consideration and its initial concentration within the droplet, we take indicative values of $\phi_{\mathrm{init}}^{*} = 10^{-6}-10^{-2}$ from \citet{Deegan2000} and indicate threshold values of $\phi_{c}^{*}/\phi_{\mathrm{init}}^{*}$ at increasing powers of 10 for illustrative purposes assuming that $\phi_{c}^{*} = 0.1$. In both regimes, as $\Pe$ increases, the solute concentration at the contact line rapidly grows and the corresponding time interval for which it is below each of the threshold values shrinks. Therefore, as $\Pe$ increases, our analysis holds for shorter periods of time. However, in the kinetic regime, even for moderately large values of the P\'{e}clet number, the solute remains dilute even at the contact line for a substantial part of the evaporative process. This suggests that there is a significant window over which the analysis presented in this paper may provide an accurate description of the early-stage formation of the coffee ring. This window grows as the P\'{e}clet number decreases or the threshold concentration $\phi_{c}^{*}/\phi_{\mathrm{init}}^{*}$ increases. Indeed, for $\Pe = 10$ and $\phi_{c}^{*}/\phi_{\mathrm{init}}^{*} = 10^{5}$, we are still in the dilute regime at $\approx 97\%$ of the drying time.

In the diffusive regime, however, things change much more rapidly because of the larger $\Pe$ scaling in (\ref{eqn:JammingDiff}). For the most moderate $\Pe$ and largest threshold $\phi_{c}^{*}/\phi_{\mathrm{init}}^{*}$ considered, we remain in the dilute regime only for $\approx 82\%$ of the drying time, which is a significant reduction from the kinetic regime. Moreover, this window of validity decreases as $\Pe$ increases and as $\phi_{c}^{*}/\phi_{\mathrm{init}}^{*}$ decreases. Even though for all cases there is a time-frame over which the dilute model presented here is relevant, it is clear that, for the case of diffusive evaporation in particular, one certainly needs to assess the importance of finite concentration effects to predict the growth and features of the coffee ring, perhaps adapting the models of \citet{Popov2005} or \citet{Kaplan2015} to include the effects of the diffusive boundary layer discussed here.

\section{Summary and discussion} 
\label{sec:Summary}

In this paper, we have presented a systematic asymptotic analysis of the solute profile as an axisymmetric droplet with a pinned contact line evaporates in the limit of large solutal P\'{e}clet-number, $\Pe$. Throughout, we have assumed that the droplet is thin and that the capillary number is small, so that surface tension dominates the droplet shape. Our analysis demonstrates that it is the effect of solute diffusion close to the contact line that can, at dilute stages of the evaporative process, drive the formation of a nascent coffee ring with its characteristic thin, narrow peak. In particular, we illustrated this behaviour for two physically-relevant evaporation models. 

Firstly, we considered the simplest kinetic regime in which the evaporative flux is constant across the drop. Even though the evaporation is uniform, the geometry of the droplet induces a radial flow of solute to the contact line, where it builds up under the effects of diffusion. Our asymptotic analysis in this boundary region produced several time-dependent coefficients that were determined by demanding that the solute is conserved within the drop (as it is non-volatile). In the kinetic regime, the boundary layer thickness is $O(R^{*}/\Pe_{t})$, the height of the solute mass profile is $O(R^{*}\Pe_{t})$ and the solute concentration is $O(\phi_{\mathrm{init}}^{*}\Pe_{t}^{2}/(1-t^{*}/t^{*}_{f}))$, where $R^{*}$ is the radius of the circular contact set, $\phi_{\mathrm{init}}^{*}$ is the initial solute concentration, $\Pe_{t} = \Pe/(1-t^{*}/t_{f}^{*})$ is the modified P\'{e}clet number, $t^{*}$ is time and $t_{f}^{*}$ is the dryout time. Asymptotic predictions of the maximum height, its radial location, the full width of the profile at half its maximum height and the composite solute mass profile were shown to be in excellent agreement with numerical solutions of the full solute problem.
% : after an initial transient in which the boundary layer forms, the maximum relative error in the asymptotic prediction for the concentration was found to be within $10\%$ for $\Pe = 10^{2}$, within $5\%$ for $\Pe = 10^{3}$ and within $2\%$ for $\Pe = 10^{4}$.

Our second example considered an evaporative model in which the liquid vapour is carried away from the droplet-air interface under the effects of diffusion. In this regime, the evaporative flux is singular at the contact line, which in turn induces a singular radial velocity. These effects combine to produce, sharper, thinner solute profiles: with a diffusive flux, the boundary layer thickness is $O(R^{*}/\Pe_{t}^{2})$, the height of the solute mass profile is $O(R^{*}\Pe_{t}^{2})$ and the solute concentration is $O(\phi_{\mathrm{init}}^{*}\Pe_{t}^{4}/(1-t^{*}/t^{*}_{f}))$, with the corresponding asymptotic predictions again shown to be in very good agreement with numerical simulations.

We then moved on to investigate the limitations of the dilute regime by using the leading-order-inner asymptotic predictions to track the growth of the solute concentration at the contact line, where it is maximal. We showed that, in each regime, there is a window in which the concentration remains below a critical value at which the effects of the finite solute particle size become important. For a fixed P\'{e}clet number, this window is longer for kinetic evaporation: indeed for $\Pe = 10$, small seeding concentration, $\phi_{\mathrm{init}}^{*} = 10^{-6}$ and threshold value $\phi_{c}^{*} = 0.1$, the dilute regime is still valid at $97\%$ of the drying time, while this falls to only $3\%$ of the drying time for $\Pe = 10^{3}$ and $\phi_{\mathrm{init}}^{*} = 10^{-2}$. In the case of diffusive evaporation, while for moderate $\Pe = 10$ and a small seeding concentration, $\phi_{\mathrm{init}}^{*} = 10^{-6}$, we remain in the dilute regime at $82\%$ of the drying time, as $\Pe$ increases and the seeding concentration increases this window diminishes quickly. Indeed, for $\Pe = 10$ and $\phi_{\mathrm{init}}^{*} = 10^{-2}$, the dilute window is only up to $4\%$ of the drying time.

These results mean that it is quite challenging to compare our theoretical and numerical results to existing experimental data for the diffusive evaporative model. Since our results concern the early stages of the development of the coffee ring, we require mass profiles long before the final deposit, which is by far the most commonly reported in existing experimental studies. However, even in cases for which transient mass profiles are given --- for example \citet{Deegan2000}, \citet{Kajiya2008} and \citet{Kajiya2011} --- the limited range of applicability for the dilute regime makes comparisons unfeasible.

\begin{figure}
\centering \scalebox{0.35}{\epsfig{file=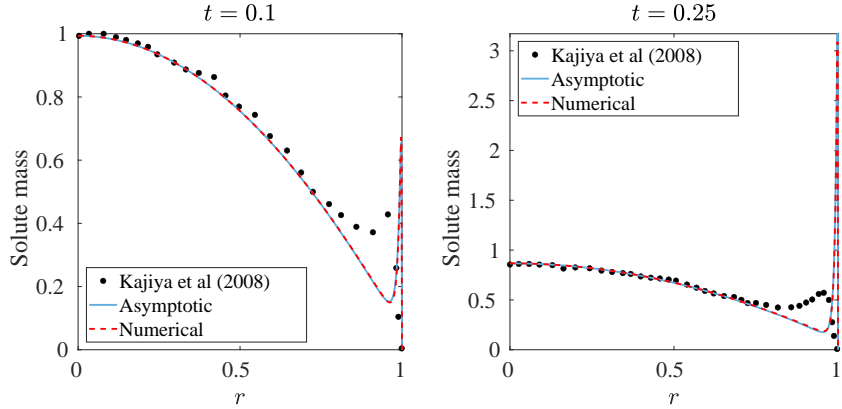}}
\caption{A comparison between the solute mass profiles in \citet{Kajiya2008} (black circles) and the asymptotic (solid, blue line) and numerical (dashed, red line) predictions of the axisymmetric model in the diffusive evaporative flux regime at 10\% (left) and 25\% (right) of the drying time. In each case, we can see that the model does well in capturing the profile towards the centre of the drop, but towards the contact line, the coffee ring is significantly thicker than the model predicts. This is due to the effect of finite particle size and solute jamming becoming important at very early times in the process as discussed in \textsection \ref{sec:Jamming}.}
\label{fig:KajiyaComparisons} 
\end{figure}

To take an example, Figure 2 of \citet{Kajiya2008} shows intensity profiles for an evaporating anisole droplet containing fluorescent polystyrene, with $\Pe\approx40$ and $\phi_{\mathrm{init}} = 0.02$. In figure \ref{fig:KajiyaComparisons}, we compare the asymptotic prediction (\ref{eqn:MCompDiff}) and the numerical results to the experimental data at 10\% and 25\% of the drying time. In the bulk of the droplet, the profiles agree extremely well, but it is clear that, as we approach the contact line, the experimental data suggests the coffee ring is much thicker than the predictions of the model. This is for precisely the reasons discussed in \textsection \ref{sec:Jamming}: for the above parameters, the dilute model breaks down when $t^{*}/t^{*}_{f} \approx 10^{-4}-10^{-3}$. Hence, our analysis provides clear evidence that the dilute model in the diffusive evaporative regime suffers a significant deficiency by discounting finite particle size effects close to the contact line.

On the other hand, as demonstrated by the large window of applicability of the dilute assumption, the kinetic evaporative problem is much riper for experimental comparison. However, much of the existing experimental literature has focused on the diffusive evaporative regime and, as far as we are aware, there is no experimental data for transient mass profiles with kinetic evaporation. It would be of great interest to conduct such comparisons between our predictions and experimental data in the future.

Our asymptotic analysis is readily extended to other geometries or physical situations. One interesting avenue for future investigation is to derive asymptotic predictions for the effect of variations in contact line curvature on the coffee ring structure. \citet{Saenz2017} showed experimentally and numerically that the coffee ring is enhanced at more highly-curved parts of the contact line and it would be interesting to predict this formally in the large-solutal P\'{e}clet number regime. While for general droplet geometries much of the problem must be tackled numerically --- particularly the Poisson problem for the free surface and finding the induced liquid velocity --- the local analysis is tractable provided that the contact line geometry is sufficiently smooth.

It would also be of interest to consider the effect of two or more droplets evaporating simultaneously, in which there is known to be a shielding effect that reduces the evaporative flux for the parts of each droplet that are closest to each other \cite[][]{Castanet2016, Hu2017}. This reduction in evaporative flux will lower the droplet flow velocities and hence weaken the coffee ring effect, and our asymptotic analysis lends itself well to predicting quantitatively how this weakening manifests itself. 

Finally, there has been a great deal of recent interest into the evaporation of binary droplets, that is droplets consisting of more than one fluid, and the resulting effects on the deposition pattern, see for example, \citet{Kim2016, Zhong2016, Li2018}. The dynamics of evaporating binary droplets is more complicated than the single liquid case presented here, with, for example, Marangoni effects becoming important. Nevertheless, it would be interesting to adapt our model for solute diffusion to investigate the role of multiple liquids.

\textbf{Acknowledgments} The authors would like to thank the anonymous referees whose comments helped improve the submitted version of this manuscript.

\textbf{Declaration of Interests.} The authors report no conflict of interests.

%%%%%%%%%%%%%%%%%%%%%%%%%%%%%%%%%%%%%%%%%%%%%%%%%%%%%%%%%%%%%%%
%%%%%%%%%%%%%%%%%%%%%%%%%% APPENDICES %%%%%%%%%%%%%%%%%%%%%%%%%
%%%%%%%%%%%%%%%%%%%%%%%%%%%%%%%%%%%%%%%%%%%%%%%%%%%%%%%%%%%%%%%

\appendix

\section{Exploiting global mass conservation}
\label{sec:ConSolAlg}

To determine the unknown functions $A(t)$, $B(t)$ and $C_{1}(t)$ in (\ref{eqn:tM0Kin}), (\ref{eqn:tM1Kin}) and (\ref{eqn:tM2Kin}), we appeal to global conservation of solute
(\ref{eqn:ConsOfSolute}). We split the range of integration such that
\begin{linenomath}
 \begin{equation}
  \int_{0}^{1}rm(r,t)\,\mbox{d}r = I_{1} + I_{2}, \; I_{1} = \int_{0}^{1-\xi}rm(r,t)\,\mbox{d}r, \; I_{2} = \int_{1-\xi}^{1} rm(r,t),\mbox{d}r, \label{eqn:CoSKin}
 \end{equation}
\end{linenomath}
where $0<\ve\ll\xi\ll1$.

In $I_{1}$, we substitute the expansion $m = m_{0}(r,t) + \ve m_{1}(r,t) + o(\ve)$ and integrate to obtain
\begin{linenomath}
\begin{eqnarray}
 I_{1} & = & \frac{a(t)}{2}\left(1-\frac{a(t)}{2}\right) + \ve\left[a(t)\left((1-a(t))t - \frac{2}{3}\left(1-a(t)^{3}\right) - \right.\right. \nonumber \\
 & & \left. \left.  2a(t)\left(1-a(t)\right)\left(a(t) - 1 + a(t)\log\left(\frac{2a(t)\xi}{1-a(t)}\right)\right)\right)\right] + o(\ve,\xi).
 \label{eqn:I1}
\end{eqnarray}
\end{linenomath}
In $I_{2}$, we make the rescaling $r = 1-\ve R$, and then expand $m = \ve^{-1} M_{0} + \log\ve M_{1} + M_{2} + o(1)$, so that after integrating and expanding as $\xi/\ve\rightarrow\infty$, we find
\begin{linenomath}
\begin{eqnarray}
 I_{2} & = & \frac{A(t)}{\alpha(t)^{2}} + \ve\log\ve \frac{(B(t)-C_{4}(t))}{\alpha(t)^{2}} + \ve\left(\frac{2(C_{2}(t)-A(t))}{\alpha(t)^{3}} + \frac{C_{1}(t)}{\alpha(t)^{2}} + \frac{6C_{3}(t)}{\alpha(t)^{4}} + \right. \nonumber \\
 & & \left. \frac{2C_{4}(t)(\gamma-1)}{\alpha(t)^{2}} + \frac{C_{4}(t)}{\alpha(t)^{2}}\log\alpha(t) + \frac{C_{4}(t)}{\alpha(t)^{2}}\log\left(\xi\right)\right) + o(\ve,\xi),
 \label{eqn:I2}
\end{eqnarray}
\end{linenomath}
where $\gamma$ is the Euler-Mascheroni constant. Then, combining (\ref{eqn:ConsOfSolute}), (\ref{eqn:I1}) and (\ref{eqn:I2}), we arrive at (\ref{eqn:beta_and_A}) and (\ref{eqn:C1}).

In the diffusive regime, the analysis is very similar, except that we must now demand $0<\ve^{2}\ll\xi\ll1$, due to the different boundary layer scalings for this evaporative flux. In this case, we have
\begin{linenomath}
 \begin{equation}
  I_{1} = \frac{1}{4}\left[1 - \left(1-(1-t)^{3/4}\right)\right]^{4/3}, \; I_{2} = \frac{3F(t)\pi^{4}}{16}(1-t)^{4},
 \end{equation}
\end{linenomath}
which, combined with (\ref{eqn:ConsOfSolute}), leads to (\ref{eqn:beta2_and_F}).

\section{Numerical method} \label{sec:NumericalMethods}

In order to solve (\ref{eqn:MassEqn})--(\ref{eqn:MassIC}) numerically, we first define the integrated mass variable
\begin{linenomath}
 \begin{equation}
  \mcalM(r,t) = \int_{0}^{r} \bar{r}m(\bar{r},t)\,\mbox{d}\bar{r}.
  \label{eqn:IMVtoMass}
 \end{equation}
\end{linenomath}
Under this transformation, the advection-diffusion equation (\ref{eqn:MassEqn}) becomes
\begin{linenomath}
 \begin{equation}
  \frac{1}{t_{f}}\frac{\partial\mcalM}{\partial t} + \left(\bur + \ve\left(\frac{1}{r}+\frac{1}{h}\frac{\partial h}{\partial r}\right)\right)\frac{\partial\mcalM}{\partial r} - \ve\frac{\partial^{2}\mcalM}{\partial r^{2}} = 0
  \label{eqn:IMVEqn}
 \end{equation}
\end{linenomath}
for $0<r<1, t>0$. This must be solved subject to 
\refstepcounter{equation}
\begin{linenomath}
 $$
  \mcalM(0,t) = 0, \; \mathcal{G}(1,t) = \frac{1}{4} \quad \mbox{for} \quad t>0,
  \eqno{(\theequation{\mathit{a},\mathit{b}})}
  \label{eqn:IMVBC}
 $$
\end{linenomath}
where the second condition (\ref{eqn:IMVBC}b) replaces the no-flux condition (\ref{eqn:MassBC}b). The initial condition is 
\begin{linenomath}
 \begin{equation}
 \mcalM(r,0) = \frac{r^{2}}{2} - \frac{r^{4}}{4} \quad \mbox{for} \quad 0<r<1.
 \label{eqn:IMVIC}
 \end{equation}
\end{linenomath}
The formulation in terms of $\mathcal{G}$ has advantages over its counterparts for the solute concentration, $\phi$, and the solute mass, $m$, since it is mass-preserving and less singular at the contact line.

We discretize (\ref{eqn:IMVEqn}) and (\ref{eqn:IMVIC}) using central differences, with the gridpoints suitably chosen to cluster in the boundary layer close to the contact line. In particular, we use a uniform grid on the computational domain $\zeta \in [0,1]$, where
\begin{linenomath}
 \begin{equation}
  r = \frac{1-\ell^{\zeta}}{1-\ell} 
 \end{equation}
\end{linenomath}
and we set the boundary layer thickness $\ell$ to be $\ell = \ve\left(1-t_{c}\right)$  in the kinetic regime and to be $\ell = \ve^{2}\left(1-t_{c}\right)^{2}$ in the diffusive regime, where $t_{c}$ is the end time of the simulations.

\begin{figure}
\centering \scalebox{0.5}{\epsfig{file=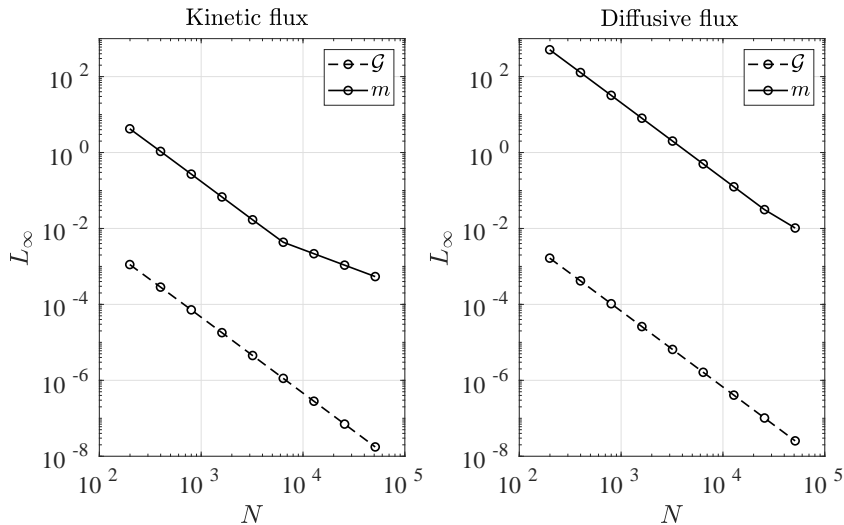}}
\caption{The $\infty$-norm error, $L_{\infty}$ in our numerical solution for the kinetic (left) and diffusive (right) evaporative fluxes as the number of gridpoints $N$ is increased. In this figure, we take $\Pe = 100$.  The dashed line in each figure denotes the error in the integrated mass variable $\mathcal{G}$, while the solid line is the resulting error in the mass, $m$.}
\label{fig:Convergence} 
\end{figure}

% \begin{figure}
% \centering \scalebox{0.5}{\epsfig{file=AsymptoticComparisons.eps}}
% \caption{The maximum relative error $|1 - \phi_{\mathrm{numer}}/\phi_{\mathrm{comp}}|$ in the solute concentration $\phi = m/h$ as a function of $t$. (Left) The kinetic model, where $\phi_{\mathrm{comp}}$ has been determined from (\ref{eqn:MCompKin}) with $\Pe = 10^{2}$ (circles), $\Pe = 10^{3}$ (triangles) and $\Pe = 10^{4}$ (squares). (Right) The diffusive model, where $\phi_{\mathrm{comp}}$ is determined from (\ref{eqn:MCompDiff}) with $\Pe = 50$ (circles), $\Pe = 100$ (triangles), $\Pe = 200$ (squares). We plot the error at time intervals of 0.1 up to $t = 0.9$.}
% \label{fig:Convergence2} 
% \end{figure}

The resulting system is solved using \emph{ode115s} in MATLAB with stringent error tolerances of $10^{-12}$ and using complex step differentiation to compute the Jacobian \cite[][]{Shampine2007}. On a standard office computer, the code runs in a few minutes for the largest numbers of gridpoints considered here (of the order of $10^{5}$). The solute mass, $m$, is recovered by differentiating (\ref{eqn:IMVtoMass}). The standard convergence checks have been performed, with an example illustrated in figure \ref{fig:Convergence} for $\Pe = 100$. We clearly see the expected quadratic convergence in $\mathcal{G}$ as the number of gridpoints, $N$, is increased. Note that the sensitivities in the diffusive code are much starker due to the extremely thin boundary layer in this regime. Nonetheless, this convergence gives us confidence in our numerical scheme, which is strengthened further by the excellent agreement with our asymptotic predictions illustrated in figures \ref{fig:ProfileComparisons}--\ref{fig:Jamming}.

% Having established convergence, we can investigate the accuracy of our asymptotic predictions by considering the maximum relative error in $\phi = m/h$ for different P\'{e}clet numbers as a function of time. We use $\phi$ rather than $m$ because $\phi$ is finite at the contact line while $m$ vanishes. We display this comparison for both evaporation models in figure \ref{fig:Convergence2}. For the kinetic evaporative model, it is clear that, after an initial transient in which the boundary layer structure forms, the relative error is a few percent for $\Pe = {10}^{2}$ and less than $1\%$ for $\Pe = 10^{3}$ and $10^{4}$. In the diffusive model, we see similar behaviour: the maximum relative error decreases for each $\Pe$, and in particular is within $5\%$ for $\Pe = 200$ --- these represent excellent accuracy for the leading-order asymptotics.

%%%%%%%%%%%%%%%%%%%%%%%%%%%%%%%%%%%%%%%%%%%%%%%%%%%%%%%%%%%%%%%
%%%%%%%%%%%%%%%%%%%%%%%%% BIBLIOGRAPHY %%%%%%%%%%%%%%%%%%%%%%%%
%%%%%%%%%%%%%%%%%%%%%%%%%%%%%%%%%%%%%%%%%%%%%%%%%%%%%%%%%%%%%%%

\bibliographystyle{jfm}


\bibliography{CoffeeRings}

\end{document}